\documentclass[a4paper,papersize,11pt]{article}
\usepackage[hidelinks]{hyperref}%
\usepackage{amsmath,amsfonts,amssymb}
\usepackage{bm}
\newcommand*\Laplace{\mathop{}\!\mathbin\bigtriangleup}
\newcommand*\DAlambert{\mathop{}\!\mathbin\Box}
\newcommand*\TimeLaplace{\mathop{}\!\mathbin{\partial_{t}^{2}}}
\newcommand{\abs}[1]{\left\lvert #1\right\rvert}
\numberwithin{equation}{section}
\def\hcentering#1{%
        \oddsidemargin=#1
        \advance\oddsidemargin-\textwidth
        \oddsidemargin=0.5\oddsidemargin
        \advance\oddsidemargin-1truein
        \evensidemargin=\oddsidemargin}
\def\vcentering#1{%
        \topmargin=#1
        \addtolength{\topmargin}{-\textheight}
        \topmargin=0.5\topmargin
        \addtolength{\topmargin}{-\headheight}
        \addtolength{\topmargin}{-\headsep}
        \footskip=\topmargin
        \addtolength{\topmargin}{-1truein}
        }

\title{Euclidean path integral of the scalar Lee-Wick model}
\author{Seiji Sakoda\thanks{E-mail: sakoda@nda.ac.jp} and Kohei Suzuki\thanks{E-mail: em57042@nda.ac.jp}\\
Department of Applied Physics, National Defense Academy, Hashirimizu\\
Yokosuka city, Kanagawa 239-8686, Japan
}

\date{\today}
\hcentering{210mm}
\vcentering{297mm}

\begin{document}
\maketitle
\abstract{
On the basis of the canonical quantization procedure, in which we need the indefinite metric Hilbert space, we formulate field diagonal representation for the scalar Lee-Wick model. Euclidean path integral for the model is then constructed in terms of  the eigenvector of field operators. Taking the quantum mechanical Lee-Wick model as an example, we demonstrate how to formulate path integrals for such systems in detail. We show that, despite the use of indefinite metric representation, integration contours for ghost degrees can be taken along the real axis.}

\section{Introduction}
Path integral is the fundamental tool for studying quantum physics. In particular, for the study of quantum field theory, its relativistic covariance gives us a useful and a powerful technique for perturbative expansion as well as non-perturbative methods such as the instanton calculation. Euclidean technique will be the foundation of such non-perturbative methods.
Formulation of the Euclidean path integral for systems quantized on the indefinite metric Hilbert space seems to remain some room to be filled by considering the construction of eigenvectors of field operators. In a recent paper\cite{Sakoda:2018Dec26}, by one of the authors (S. S.) a method has been developed to find eigenvectors of field operators for the massive vector field in which a canonical pair of variables requires quantization with the indefinite metric. His method generalizes the construction of path integrals by Arisue, Fujiwara, Inoue and Ogawa\cite{Arisue:81} and the ones by Kashiwa in ref.~\cite{Kashiwa:81} and Kashiwa and Sakamoto in ref.~\cite{Kashiwa-Sakamoto}. In this paper we shall extend this method to construct eigenvectors of field operators in the scalar Lee-Wick (LW) model\cite{Lee-Wick:NPB9,Lee-Wick:PRD2}.

The LW models, such as the Lee-Wick standard model (LWSM)\cite{LWSM}, seem recently to be intuitively studied to describe some regularized theory by removing divergent loop corrections and reconcile unitarity and renormalizability. Some phenomenological aspects of the LWSM were also reported by several authors\cite{Rizzo,Espinosa-Grinstein-O’Connell-Wise,Grinstein,Dulaney-Wise,Krauss-Underwood-Zwicky}. Path integral formulation is, however, seems to be missing in the literature for LW models. We can find a study\cite{Boulware-Gross} by Boulware and Gross on the path integral of the vector LW model. In their treatment in ref.~\cite{Boulware-Gross}, Boulware and Gross assume the positive definite representation for the original degrees and the indefinite one for ghosts. It cannot be the correct prescription to formulate Euclidean path integral for vector fields since even in the original degrees there arises the need of indefinite metric representation as has been shown in ref.~\cite{Sakoda:2018Dec26}. The propagators in the LW models will be well-defined by means of the Euclidean path integral. As was shown by Anselmi and Piva\cite{Anselmi-Piva}, propagators for LW models in Minkowski space should be defined by their Euclidean counterparts. Then we can construct Feynman rules for perturbative calculation in Minkowski space by the Wick rotation $x_{4}\to ix_{0}$ in the Euclidean propagators. In this paper we shall try to construct Euclidean path integral for the scalar LW model on a lattice in a rigorous manner by paying attention to the indefinite metric representation for the canonical commutation relations. Construction of Euclidean path integral for the vector LW model may require another kind of eigenvectors for field operators compared to the scalar LW model. We therefore discuss them separatory and restrict ourselves to the scalar LW model in this paper. Some field theoretical aspects, such as the Lorentz non-invariance\cite{Nakanishi:PRD3,Nakanishi:PRD5}, of the LW models were pointed-out and reported extensively by Nakanishi in ref.~\cite{Nakanishi:PTPS}. We may therefore do not need to reconsider such aspects in this paper again. We just aim to formulate the Euclidean path integral for the scalar LW model and give the well-defined propagator for the theory.

This paper is organized as follows: in the next section we show a detailed explanation of the method to obtain eigenvectors of field operators by making considerations on a quantum mechanical toy model. Our method given there is the extension and the refinement for the one by Aso in ref.~\cite{Aso}. The method will then be extended to the scalar LW model in section 3. We will give the Euclidean propagator there and show some calculations which clearly exhibit the finiteness of the theory. The final section will be devoted to the conclusion.
We will give an explanation for some delicate aspects in evaluating the functional determinant in the appendix.

\section{Quantum Mechanical Lee-Wick model}
Consider the Lagrangian
\begin{equation}
	L_{0}=-\frac{1}{2}q\left(\partial_{t}^{2}+m^{2}\right)
	\left(1+\frac{\partial_{t}^{4}}{M^{4}}\right)q.
\end{equation}
To quantize and formulate a path integral for systems of this kind requires a prescription how to introduce canonical pairs for which canonical commutation relations are defined. On this issue, a very clear explanation by Nakanishi is found in ref.~\cite{Nakanishi:PTPS}. According to Nakanishi's prescription, we consider a polynomial $F(s)$ given by
\begin{equation}
	F(s)=(s-s_{0})(s-s_{+})(s-s_{-}),
\end{equation}
where $s_{0}=-m^{2}$ and $s_{\pm}=\mp iM^{2}$.
Since there holds
\begin{equation}
	\frac{1}{F(s)}=
	\sum_{j=0,\pm}\frac{1}{F{'}(s_{j})(s-s_{j})},
\end{equation}
we obtain
\begin{equation}
	1=F(\TimeLaplace)\sum_{j=0,\pm}\frac{1}{F{'}(s_{j})(\TimeLaplace-s_{j})}
\end{equation}
to find
\begin{equation}
	q=\tilde{q}_{0}+\tilde{\phi}+\tilde{\phi}^{\dagger},
\end{equation}
where
\begin{equation}
	\tilde{q}_{0}=\frac{1}{F{'}(s_{0})}
	\left(\TimeLaplace+iM^{2}\right)\left(\TimeLaplace-iM^{2}\right)q,
\end{equation}
and
\begin{equation}
	\tilde{\phi}=\frac{1}{F{'}(s_{+})}
	\left(\TimeLaplace+m^{2}\right)\left(\TimeLaplace-iM^{2}\right)q,\quad
	\tilde{\phi}^{\dagger}=\frac{1}{F{'}(s_{-})}
	\left(\TimeLaplace+m^{2}\right)\left(\TimeLaplace+iM^{2}\right)q.
\end{equation}
We can then rewrite the Lagrangian as
\begin{equation}
	L_{0}=
	\frac{\abs{W}^{2}}{2}
	\left(\dot{\tilde{q}}_{0}^{2}-m^{2}\tilde{q}_{0}^{2}\right)
	-W\left(\dot{\tilde{\phi}}^{2}-iM^{2}\tilde{\phi}^{2}\right)
	-W^{*}\left(
	\dot{\tilde{\phi}}^{\dagger\,2}+iM^{2}\tilde{\phi}^{\dagger\,2}\right)
\end{equation}
where we have written
\begin{equation}
	\frac{F{'}(s_{0})}{M^{4}}=1+\frac{m^{4}}{M^{4}}=\abs{W}^{2},\quad
	\frac{F{'}(s_{+})}{M^{4}}=-2\left(1+i\frac{m^{2}}{M^{2}}\right)=
	-2W.
\end{equation}
To simplify the notation and the formulation below, we redefine variables as
\begin{equation}
	q_{0}\equiv\abs{W}\tilde{q}_{0},\quad
	\phi\equiv i\sqrt{2W}\tilde{\phi},\quad
	\phi^{\dagger}\equiv-i\sqrt{2W^{*}}\tilde{\phi}^{\dagger}.
\end{equation}
We then obtain
\begin{equation}
	L_{0}=\frac{1}{2}
	\left(\dot{q}_{0}^{2}-m^{2}q_{0}^{2}\right)
	+\frac{1}{2}\left(\dot{\phi}^{2}-iM^{2}\phi^{2}\right)
	+\frac{1}{2}\left(
	\dot{\phi}^{\dagger\,2}+iM^{2}\phi^{\dagger\,2}\right).
\end{equation}

Canonical conjugates for $q_{0}$, $\phi$ and $\phi^{\dagger}$ are given by
\begin{equation}
	p_{0}=\dot{q}_{0},\quad
	\pi=\dot{\phi},\quad
	\pi^{\dagger}=\dot{\phi}^{\dagger},
\end{equation}
respectively and the Hamiltonian reads
\begin{equation}
	H_{0}=\frac{1}{2}\left(p_{0}^{2}+m^{2}q_{0}^{2}\right)+
	\frac{1}{2}\left(\pi^{2}+\omega^{2}\phi^{2}\right)+
	\frac{1}{2}\left(\pi^{\dagger\,2}+\omega^{*\,2}\phi^{\dagger\,2}\right),
	\quad
	\omega^{2}\equiv iM^{2}.
\end{equation}
Despite the appearance of $\pm iM^{2}$ in the second and the third term, the Hamiltonian is Hermitian at least formally. We therefore take this Hamiltonian as the free Hamiltonian of the quantum mechanics of this system and denote operators by $\hat{p}_{0}$, $\hat{q}_{0}$, etc. hereafter.
As for the degree described by $\hat{q}_{0}$, the system is nothing but a usual harmonic oscillator whose quantum behavior is quite wellknown.
We can thus achieve the quantization procedure for this degree by the use of ordinary Schr\"{o}dinger representation of the canonical commutation relation
\begin{equation}
	[\hat{q}_{0},\,\hat{p}_{0}]=i.
\end{equation}
We may define the creation and annihilation operators for this degree by
\begin{equation}
	\hat{a}_{0}=\sqrt{\frac{m}{2}}\left(\hat{q}_{0}+
	\frac{i}{m}\hat{p}_{0}\right),\quad
	\hat{a}_{0}^{\dagger}=\sqrt{\frac{m}{2}}\left(\hat{q}_{0}-
	\frac{i}{m}\hat{p}_{0}\right),
\end{equation}
to describe the Fock representation for the quantum mechanics of this oscillator.

On the other hand, the corresponding ones for the degrees of $\hat{\phi}$ and $\hat{\phi}^{\dagger}$:
\begin{equation}
	[\hat{\phi}_{0},\,\hat{\pi}_{0}]=
	[\hat{\phi}^{\dagger}_{0},\,\hat{\pi}^{\dagger}_{0}]=i
\end{equation}
lead to non-standard creation and annihilation operators through 
\begin{equation}
\begin{gathered}
	\hat{\phi}(t)=\frac{1}{\sqrt{2\omega}}\left(
	\hat{\alpha}e^{-i\omega t}+\hat{\beta}^{\dagger}e^{i\omega t}
	\right),\quad
	\hat{\phi}^{\dagger}(t)=\frac{1}{\sqrt{2\omega^{*}}}\left(
	\hat{\alpha}^{\dagger}e^{i\omega^{*}t}+
	\hat{\beta}e^{-i\omega^{*}t}
	\right),\\
	\hat{\pi}(t)=-i\sqrt{\frac{\omega}{2}}\left(
	\hat{\alpha}e^{-i\omega t}-\hat{\beta}^{\dagger}e^{i\omega t}
	\right),\quad
	\hat{\pi}^{\dagger}(t)=i\sqrt{\frac{\omega^{*}}{2}}\left(
	\hat{\alpha}^{\dagger}e^{i\omega^{*}t}-
	\hat{\beta}e^{-i\omega^{*}t}
	\right),
\end{gathered}
\end{equation}
in which $\hat{\alpha}$, $\hat{\beta}$, $\hat{\alpha}^{\dagger}$ and $\hat{\beta}^{\dagger}$ satisfy
\begin{equation}
	[\hat{\alpha},\,\hat{\beta}^{\dagger}]=
	[\hat{\beta},\,\hat{\alpha}^{\dagger}]=1
\end{equation}
in addition to the other trivial ones.

We can assume the existence of the eigenvectors of $\hat{q}_{0}$ and $\hat{p}_{0}$ to satisfy
\begin{equation}
	\hat{q}_{0}\vert q_{0}\rangle=\vert q_{0}\rangle q_{0},\quad
	\hat{p}_{0}\vert p_{0}\rangle=\vert p_{0}\rangle p_{0}
\end{equation}
in addition to their Hermitian conjugates as the fundamental ingredient in construction of the path integral. We here rather proceed in a different way to find the explicit form of these eigenvectors in terms of the state vectors in the Fock space. To this aim let us consider
\begin{equation}
	\hat{\Pi}_{0}(q_{0})=\frac{1}{2\pi}\int_{-\infty}^{\infty}\!\!d\lambda
	\exp\left\{i\lambda\left(\hat{q}_{0}-q_{0}\right)\right\}.
\end{equation}
By formal manipulation, we find
\begin{equation}
	\int_{-\infty}^{\infty}\!\!\hat{\Pi}_{0}(q_{0})\,dq_{0}=1.
\end{equation}
Hence we can expect $\hat{\Pi}_{0}(q_{0})$ defines the projection
\begin{equation}
	\hat{\Pi}_{0}(q_{0})=\vert q_{0}\rangle\langle q_{0}\vert.
\end{equation}
In order to confirm this, we first define the coherent state $\vert z\rangle$ by
\begin{equation}
	\vert z\rangle=e^{\hat{a}_{0}^{\dagger}z}\vert0\rangle,
\end{equation}
where the vacuum $\vert0\rangle$ is defined to satisfy
\begin{equation}
	\hat{a}_{0}\vert0\rangle=
	\hat{\alpha}\vert0\rangle=
	\hat{\beta}\vert0\rangle=0,\quad
	\langle0\vert0\rangle=1.
\end{equation}
We then sandwich $\hat{\Pi}_{0}(q_{0})$ above by $\langle z\vert$ and $\vert z{'}\rangle$ to find
\begin{equation}
	\langle z\vert\hat{\Pi}_{0}(q_{0})\vert z{'}\rangle=
	\frac{1}{2\pi}\int_{-\infty}^{\infty}\!\!d\lambda\,
	\exp\left\{-\frac{\lambda^{2}}{4m}+i\lambda\left(
	q_{0}^{(+)}+q_{0}^{(-)}-q_{0}\right)\right\}
	\langle z\vert z{'}\rangle,
\end{equation}
where
\begin{equation}
	q_{0}^{(+)}=\frac{z{'}}{\sqrt{2m}},\quad
	q_{0}^{(-)}=\frac{z^{*}}{\sqrt{2m}}
\end{equation}
and the inner product of coherent states is given by $\langle z\vert z{'}\rangle=e^{z^{*}z{'}}$.
By carrying out the Gaussian integral, we obtain
\begin{equation}
	\langle z\vert\hat{\Pi}_{0}(q_{0})\vert z{'}\rangle=\sqrt{\frac{m}{\pi}}
	\exp\left\{-m\left(q_{0}^{(+)}+q_{0}^{(-)}-q_{0}\right)^{2}\right\}
	\langle z\vert z{'}\rangle.
\end{equation}
On substitution of $q_{0}^{(\pm)}$ above and noticing that $e^{-z^{*}z{'}}$ cancels the inner product $\langle z\vert z{'}\rangle$, we find
\begin{equation}
	\langle z\vert\hat{\Pi}_{0}(q_{0})\vert z{'}\rangle=\sqrt{\frac{m}{\pi}}
	\exp\left\{-mq_{0}^{2}-\frac{1}{2}\left(z^{*\,2}+z{'}^{2}\right)+
	\sqrt{2m}q_{0}\left(z^{*}+z{'}\right)\right\}.
\end{equation}
Since there holds $\hat{a}_{0}\vert z{'}\rangle=\vert z{'}\rangle z{'}$ as well as $\langle z\vert\hat{a}_{0}^{\dagger}=z^{*}\langle z\vert$ in addition to $\langle0\vert z{'}\rangle=\langle z\vert0\rangle=1$, we can rewrite it as
\begin{equation}
	\langle z\vert\hat{\Pi}_{0}(q_{0})\vert z{'}\rangle=\sqrt{\frac{m}{\pi}}
	e^{-mq_{0}^{2}}\langle z\vert
	e^{-\hat{a}_{0}^{\dagger\,2}/2+\sqrt{2m}q_{0}\hat{a}_{0}^{\dagger}}
	\vert0\rangle\langle0\vert
	e^{-\hat{a}_{0}^{2}/2+\sqrt{2m}q_{0}\hat{a}_{0}}\vert z{'}\rangle
\end{equation}
which clearly shows the correctness of our prospect above. We thus obtain
\begin{equation}
\begin{aligned}
	\vert q_{0}\rangle=&\left(\frac{m}{\pi}\right)^{1/4}
	\exp\left(-\frac{m}{2}q_{0}^{2}-\frac{1}{2}\hat{a}_{0}^{\dagger\,2}+
	\sqrt{2m}q_{0}\hat{a}_{0}^{\dagger}\right)
	\vert0\rangle,\\
	\langle q_{0}\vert=&\left(\frac{m}{\pi}\right)^{1/4}\langle0\vert
	\exp\left(-\frac{m}{2}q_{0}^{2}-\frac{1}{2}\hat{a}_{0}^{2}+
	\sqrt{2m}q_{0}\hat{a}_{0}\right).
\end{aligned}
\end{equation}
In the same way, we can find eigenvectors of the momentum operator $\hat{p}_{0}$ given by
\begin{equation}
\begin{aligned}
	\vert p_{0}\rangle=&\frac{1}{(m\pi)^{1/4}}
	\exp\left(-\frac{1}{2m}p_{0}^{2}+\frac{1}{2}\hat{a}_{0}^{\dagger\,2}+
	i\sqrt{\frac{2}{m}}p_{0}\hat{a}_{0}^{\dagger}\right)
	\vert0\rangle,\\
	\langle p_{0}\vert=&\frac{1}{(m\pi)^{1/4}}\langle0\vert
	\exp\left(-\frac{1}{2m}p_{0}^{2}+\frac{1}{2}\hat{a}_{0}^{2}-
	i\sqrt{\frac{2}{m}}p_{0}\hat{a}_{0}\right).
\end{aligned}
\end{equation}
In the calculation of inner products of these eigenvectors, it is convenient to linearize $\hat{a}_{0}^{\dagger\,2}$ and $\hat{a}_{0}^{2}$ by making use of integral representations, such as
\begin{equation}
	e^{-\hat{a}_{0}^{\dagger\,2}}=\frac{1}{\sqrt{2\pi}}
	\int_{-\infty}^{\infty}\!\!du\,
	e^{-u^{2}/2+i\hat{a}_{0}^{\dagger}u}.
\end{equation}
This is equivalent to view eigenvectors above as Gaussian averages of the coherent states. It is then straightforward to see
\begin{equation}
	\langle q_{0}\vert q_{0}{'}\rangle=\delta(q_{0}-q_{0}{'}),\quad
	\langle p_{0}\vert p_{0}{'}\rangle=\delta(p_{0}-p_{0}{'}),\quad
	\langle q_{0}\vert p_{0}\rangle=\langle p_{0}\vert q_{0}\rangle^{*}=
	\frac{1}{\sqrt{2\pi}}e^{iqp}.
\end{equation}
As already pointed out above, our method of finding eigenvectors includes inside itself the explicit form of the resolutions of unity expressed in terms of eigenvectors to be obtained. We are therefore ready to formulate path integrals so long as the $\hat{q}_{0}$ degree of freedom is concerned.

To achieve the set up for other degrees, we have to look carefully the structure of the representation space for the algebra of $\hat{\alpha}$, $\hat{\beta}$, $\hat{\alpha}^{\dagger}$ and $\hat{\beta}^{\dagger}$. If we define, for integers $n_{1}$ and $n_{2}$,
\begin{equation}
	\vert n_{1},\,n_{2}\rangle=\frac{1}{\sqrt{n_{1}!n_{2}!}}
	\hat{\alpha}^{\dagger\,n_{1}}\hat{\beta}^{\dagger\,n_{2}}\vert0\rangle,
\end{equation}
we observe
\begin{equation}
	\langle n_{1},\,n_{2}\vert n{'}_{1},\,n{'}_{2}\rangle=
	\delta_{n_{1},\,n{'}_{2}}\delta_{n_{2},\,n{'}_{1}},
\end{equation}
where $\langle n_{1},\,n_{2}\vert=\vert n_{1},\,n_{2}\rangle^{\dagger}$.
We therefore need $\langle\underline{n_{1},\,n_{2}}\vert=\vert n_{2},\,n_{1}\rangle^{\dagger}$ as the conjugate of $\vert n_{1},\,n_{2}\rangle$ to write the resolution of unity for these degrees of freedom as
\begin{equation}
	\sum_{n_{1},\,n_{2}=0}^{\infty}\vert n_{1},\,n_{2}\rangle
	\langle\underline{n_{1},\,n_{2}}\vert=1.
\end{equation}
Basing on these observations, we define a coherent state by
\begin{equation}
	\vert\alpha,\,\beta\rangle=
	e^{\alpha\hat{\beta}^{\dagger}+\beta\hat{\alpha}^{\dagger}}\vert0\rangle
\end{equation}
to fulfill
\begin{equation}
	\hat{\alpha}\vert\alpha,\,\beta\rangle=
	\vert\alpha,\,\beta\rangle\alpha,\quad
	\hat{\beta}\vert\alpha,\,\beta\rangle=
	\vert\alpha,\,\beta\rangle\beta.
\end{equation}
Taking the metric structure above into account, $\langle\underline{\alpha,\,\beta}\vert$ is defined as the conjugate to $\vert\alpha,\,\beta\rangle$ by
\begin{equation}
	\langle\underline{\alpha,\,\beta}\vert=\langle0\vert
	e^{\alpha^{*}\hat{\alpha}+\beta^{*}\hat{\beta}}
\end{equation}
to satisfy
\begin{equation}
	\langle\underline{\alpha,\,\beta}\vert\hat{\alpha}^{\dagger}=
	\beta^{*}\langle\underline{\alpha,\,\beta}\vert,\quad
	\langle\underline{\alpha,\,\beta}\vert\hat{\beta}^{\dagger}=
	\alpha^{*}\langle\underline{\alpha,\,\beta}\vert
\end{equation}
and yield the inner product $\langle\underline{\alpha,\,\beta}\vert\alpha{'},\,\beta{'}\rangle=e^{\alpha^{*}\alpha{'}+\beta^{*}\beta{'}}$. It is easy to prove the resolution of unity
\begin{equation}
	\int\!\!\frac{d\alpha\,d\alpha^{*}\,d\beta\,d\beta^{*}}{\pi^{2}}
	e^{-\alpha^{*}\alpha-\beta^{*}\beta}
	\vert\alpha,\,\beta\rangle\langle\underline{\alpha,\,\beta}\vert=1,
\end{equation}
where integrations should be taken with respect to real parts and imaginary parts of $\alpha$ and $\beta$.

We now consider
\begin{equation}
	\hat{\Pi}(\phi,\,\phi^{\dagger})=\frac{1}{(2\pi)^{2}}
	\int_{\varGamma_{\lambda},\,\varGamma_{\lambda^{*}}}\!\!
	d\lambda\,d\lambda^{*}
	\exp\left\{i\lambda^{*}(\hat{\phi}-\phi)+
	i\lambda(\hat{\phi}^{\dagger}-\phi^{\dagger})\right\}
\end{equation}
and sandwich it by $\langle\underline{\alpha,\,\beta}\vert$ and $\vert\alpha{'},\,\beta{'}\rangle$. Here, as will be made clear soon below, the contours of integrations above are taken as follows: $\lambda^{*}$ ($\lambda$) runs on a line on which $\lambda^{*\,2}/\omega$ ($\lambda^{2}/\omega^{*}$) takes real values.
That there hold
\begin{equation}
	\hat{\phi}\hat{\Pi}(\phi,\,\phi^{\dagger})=
	\hat{\Pi}(\phi,\,\phi^{\dagger})\hat{\phi}=
	\phi\hat{\Pi}(\phi,\,\phi^{\dagger}),\quad
	\hat{\phi}^{\dagger}\hat{\Pi}(\phi,\,\phi^{\dagger})=
	\hat{\Pi}(\phi,\,\phi^{\dagger})\hat{\phi}^{\dagger}=
	\phi^{\dagger}\hat{\Pi}(\phi,\,\phi^{\dagger})
\end{equation}
is easy to see if we make use of the integration by parts.
By making use of the commutation relations, we easily obtain
\begin{multline}
	\hat{\Pi}(\phi,\,\phi^{\dagger})=\frac{1}{(2\pi)^{2}}
	\int_{\varGamma_{\lambda},\,\varGamma_{\lambda^{*}}}\!\!
	d\lambda\,d\lambda^{*}
	\exp\left\{-\frac{\lambda^{*\,2}}{4\omega}-
	\frac{\lambda^{2}}{4\omega^{*}}-
	i\lambda^{*}\phi-i\lambda\phi^{\dagger}\right\}\\
	\times
	\exp\left(i\lambda^{*}\hat{\phi}^{(-)}+
	i\lambda\hat{\phi}^{\dagger\,(-)}\right)
	\exp\left(i\lambda^{*}\hat{\phi}^{(+)}+
	i\lambda\hat{\phi}^{\dagger\,(+)}\right),
\end{multline}
where
\begin{equation}
	\hat{\phi}^{(+)}=\frac{\hat{\alpha}}{\sqrt{2\omega}},\quad
	\hat{\phi}^{(-)}=\frac{\hat{\beta}^{\dagger}}{\sqrt{2\omega}},\quad
	\hat{\phi}^{\dagger\,(+)}=\frac{\hat{\beta}}{\sqrt{2\omega^{*}}},\quad
	\hat{\phi}^{\dagger\,(-)}=\frac{\hat{\alpha}^{\dagger}}{\sqrt{2\omega^{*}}}
\end{equation}
and they take eigenvalues on coherent states as
\begin{equation}
\begin{gathered}
	\hat{\phi}^{(+)}\vert\alpha{'},\,\beta{'}\rangle=
	\vert\alpha{'},\,\beta{'}\rangle
	\phi^{(+)},\quad
	\hat{\phi}^{\dagger\,(+)}\vert\alpha{'},\,\beta{'}\rangle=
	\vert\alpha{'},\,\beta{'}\rangle
	\phi^{\dagger\,(+)},\\
	\langle\underline{\alpha,\,\beta}\vert\hat{\phi}^{(-)}=
	\phi^{(-)}
	\langle\underline{\alpha,\,\beta}\vert,\quad
	\langle\underline{\alpha,\,\beta}\vert\hat{\phi}^{\dagger\,(-)}=
	\phi^{\dagger\,(-)}
	\langle\underline{\alpha,\,\beta}\vert,
\end{gathered}
\end{equation}
where
\begin{equation}
	\phi^{(+)}=\frac{\alpha{'}}{\sqrt{2\omega}},\quad
	\phi^{\dagger\,(+)}=\frac{\beta{'}}{\sqrt{2\omega^{*}}},\quad
	\phi^{(-)}=\frac{\alpha^{*}}{\sqrt{2\omega}},\quad
	\phi^{\dagger\,(-)}=\frac{\beta^{*}}{\sqrt{2\omega^{*}}}.
\end{equation}
On the integration contours of $\lambda$ and $\lambda^{*}$, we define Gaussian integrals along their own contours; we can, however, modify these to the real axis because both $\omega$ and $\omega^{*}$ possess positive real parts as far as $M>0$. The same holds integrals appear below.
We thus find
\begin{equation}
	\langle\underline{\alpha,\,\beta}\vert
	\hat{\Pi}(\phi,\,\phi^{\dagger})\vert\alpha{'},\,\beta{'}\rangle=
	\frac{\abs{\omega}}{\pi}
	\exp\left\{-\omega\left(\phi-\phi^{(+)}-\phi^{(-)}\right)^{2}-
	\omega^{*}\left(
	\phi^{\dagger}-\phi^{\dagger\,(+)}-\phi^{\dagger\,(-)}
	\right)^{2}\right\}
	\langle\underline{\alpha,\,\beta}\vert\alpha{'},\,\beta{'}\rangle.
\end{equation}
By choosing the integration contours for $\phi$ ($\phi^{\dagger}$) such that $\omega\phi^{2}$ ($\omega^{*}\phi^{\dagger\,2}$) takes real values, we can carry out Gaussian integrations with respect to these variables and obatin
\begin{equation}
\label{eq:res_qm01}
	\int_{\varGamma_{\phi},\,\varGamma_{\phi^{\dagger}}}\!\!
	d\phi\,d\phi^{\dagger}\,
	\langle\underline{\alpha,\,\beta}\vert
	\hat{\Pi}(\phi,\,\phi^{\dagger})\vert\alpha{'},\,\beta{'}\rangle=
	\langle\underline{\alpha,\,\beta}\vert\alpha{'},\,\beta{'}\rangle
\end{equation}
which clearly exhibits that there holds
\begin{equation}
	\int_{\varGamma_{\phi},\,\varGamma_{\phi^{\dagger}}}\!\!
	d\phi\,d\phi^{\dagger}\,
	\hat{\Pi}(\phi,\,\phi^{\dagger})=1
\end{equation}
as a resolution of unity. We can therefore identify as
\begin{equation}
	\hat{\Pi}(\phi,\,\phi^{\dagger})=\vert\phi,\,\phi^{\dagger}\rangle
	\langle\underline{\phi,\,\phi^{\dagger}}\vert
\end{equation}
where
\begin{equation}
	\vert\phi,\,\phi^{\dagger}\rangle=\sqrt{\frac{\abs{\omega}}{\pi}}
	\exp\left\{-\frac{\omega}{2}\phi^{2}-\frac{\omega^{*}}{2}\phi^{\dagger\,2}-
	\frac{1}{2}\left(\hat{\alpha}^{\dagger\,2}+
	\hat{\beta}^{\dagger\,2}\right)-
	\sqrt{2\omega}\phi\hat{\beta}^{\dagger}-
	\sqrt{2\omega^{*}}\phi^{\dagger}\hat{\alpha}^{\dagger}\right\}\vert0\rangle
\end{equation}
and
\begin{equation}
	\langle\underline{\phi,\,\phi^{\dagger}}\vert=
	\sqrt{\frac{\abs{\omega}}{\pi}}\langle0\vert
	\exp\left\{-\frac{\omega}{2}\phi^{2}-\frac{\omega^{*}}{2}\phi^{\dagger\,2}-
	\frac{1}{2}\left(\hat{\alpha}^{2}+\hat{\beta}^{2}\right)-
	\sqrt{2\omega}\phi\hat{\alpha}-
	\sqrt{2\omega^{*}}\phi^{\dagger}\hat{\beta}\right\}.
\end{equation}
The inner product between eigenvectors is given by
\begin{equation}
	\langle\underline{\phi,\,\phi^{\dagger}}\vert
	\phi{'},\,\phi{'}^{\dagger}\rangle=
	\delta_{\varGamma_{\phi}}(\phi-\phi{'})
	\delta_{\varGamma_{\phi^{\dagger}}}(\phi^{\dagger}-\phi{'}^{\dagger})
\end{equation}
in which delta functions are complex delta functions\cite{Nakanishi:PTPS} defined on each integration contour. It should be noted here that we can modify both the integration contours $\varGamma_{\phi}$ and $\varGamma_{\phi^{\dagger}}$ to the real axis in \eqref{eq:res_qm01} thanks to the existence of positive real parts both in $\omega$ and $\omega^{*}$. In this regard, the indefinite metric representation yields imaginary axis in the Euclidean path integral of vector field. It is therefore in a quite sharp contrast that field variables of ghost degrees can be taken to be real in the Euclidean path integral of quantum mechanical LW model. We will see this is true even for the field theoretical scalar LW model in the next section.

Finding eigenvectors of momentum operators $\hat{\pi}$ and $\hat{\pi}^{\dagger}$ can be formulated in the same manner. We may not need to repeat showing the detail and will be allowed to give the results here;
\begin{equation}
\begin{aligned}
	\vert\varPi,\varPi^{\dagger}\rangle=
	&\frac{1}{\sqrt{\abs{\omega}\pi}}
	\exp\left\{-\frac{1}{2\omega}\varPi^{2}-
	\frac{1}{2\omega^{*}}\varPi^{\dagger\,2}+
	\frac{1}{2}\left(\hat{\alpha}^{\dagger\,2}+
	\hat{\beta}^{\dagger\,2}\right)-
	i\sqrt{\frac{2}{\omega}}\varPi\hat{\beta}^{\dagger}-
	i\sqrt{\frac{2}{\omega^{*}}}\varPi^{\dagger}
	\hat{\alpha}^{\dagger}\right\}\vert0\rangle,\\
	\langle\underline{\varPi,\varPi^{\dagger}}\vert=
	&\frac{1}{\sqrt{\abs{\omega}\pi}}
	\langle0\vert
	\exp\left\{-\frac{1}{2\omega}\varPi^{2}-
	\frac{1}{2\omega^{*}}\varPi^{\dagger\,2}+
	\frac{1}{2}\left(\hat{\alpha}^{2}+
	\hat{\beta}^{2}\right)+
	i\sqrt{\frac{2}{\omega}}\varPi\hat{\alpha}+
	i\sqrt{\frac{2}{\omega^{*}}}\varPi^{\dagger}
	\hat{\beta}\right\},
\end{aligned}
\end{equation}
where $\varPi$ and $\varPi^{\dagger}$ are eigenvalues of $\hat{\pi}$ and $\hat{\pi}^{\dagger}$, respectively.
Resolution of unity expressed by the momentum eigenvector reads
\begin{equation}
	\int_{\varGamma_{\pi},\,\varGamma_{\pi^{\dagger}}}\!\!
	d\varPi\,d\varPi^{\dagger}
	\vert\varPi,\varPi^{\dagger}\rangle\langle
	\underline{\varPi,\varPi^{\dagger}}\vert=1,
\end{equation}
where $\varGamma_{\pi}$ ($\varGamma_{\pi^{\dagger}}$) is the integration contour for $\varPi$ ($\varPi^{\dagger}$) on which $\varPi^{2}/\omega$ ($\varPi^{\dagger\,2}/\omega^{*}$) takes real values.
The inner product is again defined as a product of complex delta functions on each integration contour as
\begin{equation}
	\langle\underline{\varPi,\varPi^{\dagger}}\vert
	\varPi{'},\varPi{'}^{\dagger}\rangle=
	\delta_{\varGamma_{\pi}}(\varPi-\varPi{'})
	\delta_{\varGamma_{\pi^{\dagger}}}(\varPi^{\dagger}-\varPi{'}^{\dagger}),
\end{equation}
while those between a momentum eigenvector and a position eigenvector being given by
\begin{equation}
	\langle\underline{\phi,\,\phi^{\dagger}}\vert
	\varPi,\varPi^{\dagger}\rangle=
	\langle\underline{\varPi,\,\varPi^{\dagger}}\vert
	\phi,\phi^{\dagger}\rangle^{*}=
	\frac{1}{2\pi}e^{i\phi\varPi+i\phi^{\dagger}\varPi^{\dagger}}.
\end{equation}
We have thus completed the preparation for constructing path integrals for the system under consideration.

Let us now turn to formulation of path integrals. We first consider the Euclidean path integral of the generating functional $Z^{(\mathrm{E})}[J_{a}]$. As the first step for this aim, we consider a short time kernel
\begin{equation}
	K^{(\mathrm{E})}_{J}(\varPhi,\varPhi{'};\epsilon)=
	\langle\underline{\varPhi}\vert\left(1-\epsilon\hat{H}_{0}
	-\epsilon J_{a}\hat{\varPhi}_{a}\right)\vert\varPhi{'}\rangle,
\end{equation}
where $\varPhi$ designates symbolically and uniformly $q_{0}$, $\phi$ and $\phi^{\dagger}$ and in the source term $J_{a}\hat{\varPhi}_{a}$ the suffix $a$ runs over these elements. By making use of the momentum eigenvectors, we can evaluate the short time kernel above, after Gaussian integrations with respect to momentum variables, turns out to be
\begin{multline}
	K^{(\mathrm{E})}_{J}(\varPhi,\varPhi{'};\epsilon)=
	\frac{1}{(2\pi\epsilon)^{3/2}}\\
	\times
	\exp\left[-\frac{1}{2\epsilon}\left\{
	(\Delta q_{0})^{2}+(\Delta\phi)^{2}+(\Delta\phi^{\dagger})^{2}\right\}-
	\frac{\epsilon}{2}\left\{
	m^{2}q_{0}^{2}+\omega^{2}\phi^{2}+\omega^{*\,2}\phi^{\dagger\,2}\right\}-
	\epsilon\left(J_{0}q_{0}+J^{*}\phi+J\phi^{\dagger}\right)\right],
\end{multline}
where $\Delta q_{0}=q_{0}-q_{0}{'}$, $\Delta\phi=\phi-\phi{'}$ and $\Delta\phi^{\dagger}=\phi^{\dagger}-\phi{'}^{\dagger}$.
We devide the Euclidean time interval $-T/2<t<T/2$ into $2N+1$ equal length segments and require periodic boundary condition.
The generating functional will be evaluated by integrating
\begin{equation}
	Z^{(\mathrm{E})}[J]=\int\!\!\prod_{j=-N}^{N}\prod_{a}d\varPhi_{a;\,j}
	K^{(\mathrm{E})}_{J}(\varPhi_{N},\varPhi_{N-1};\epsilon)
	K^{(\mathrm{E})}_{J}(\varPhi_{N-1},\varPhi_{N-2};\epsilon)\times\cdots\times
	K^{(\mathrm{E})}_{J}(\varPhi_{-N},\varPhi_{N};\epsilon).
\end{equation}
As is evident, however, from the action of the short time kernel, this time sliced path integral is factorized into the ones for each degree. We can therefore carry out the path integral to obtain $Z^{(\mathrm{E})}[J_{a}]$ in the form  $Z^{(\mathrm{E})}[J_{a}]=Z^{(\mathrm{E})}_{q_{0}}[J_{0}]Z^{(\mathrm{E})}_{\phi}[J^{*}]Z^{(\mathrm{E})}_{\phi^{\dagger}}[J]$. Since the action is quadratic, though with complex coefficients, the evaluation of each path integral is straightforward. We therefore obtain
\begin{equation}
	Z^{(\mathrm{E})}_{q_{0}}[J_{0}]=\frac{1}{2\sinh(mT/2)}
	e^{-W^{(\mathrm{E})}_{q_{0}}[J_{0}]},\quad
	W^{(\mathrm{E})}_{q_{0}}[J_{0}]=-\frac{\epsilon^{2}}{2}\sum_{j,\,k=-N}^{N}
	J_{0;\,j}\Delta^{(m)}_{j,\,k}J_{0;\,k},
\end{equation}
\begin{equation}
	Z^{(\mathrm{E})}_{\phi}[J^{*}]=\frac{1}{2\sinh(\omega T/2)}
	e^{-W^{(\mathrm{E})}_{\phi}[J^{*}]},\quad
	W^{(\mathrm{E})}_{\phi}[J^{*}]=-\frac{\epsilon^{2}}{2}\sum_{j,\,k=-N}^{N}
	J^{*}_{j}\Delta^{(\omega)}_{j,\,k}J^{*}_{k},
\end{equation}
and
\begin{equation}
	Z^{(\mathrm{E})}_{\phi^{\dagger}}[J]=\frac{1}{2\sinh(\omega^{*}T/2)}
	e^{-W^{(\mathrm{E})}_{\phi^{\dagger}}[J]},\quad
	W^{(\mathrm{E})}_{\phi^{\dagger}}[J]=
	-\frac{\epsilon^{2}}{2}\sum_{j,\,k=-N}^{N}
	J_{j}\Delta^{(\omega^{*})}_{j,\,k}J_{k},
\end{equation}
respectively. Here the propagator for each degree is given by the substitution of $m$, $\omega$ and $\omega^{*}$ to $\mu$ in 
\begin{equation}
	\Delta^{(\mu)}_{j,\,k}=\frac{1}{\epsilon}\sum_{n=-N}^{N}
	\frac{F_{j}^{n}F_{k}^{n\,*}}
	{\dfrac{4}{\epsilon^{2}}\sin^{2}\left(\dfrac{\pi n}{2N+1}\right)+\mu^{2}}
	,\quad
	F_{j}^{n}=\frac{1}{\sqrt{2N+1}}e^{2\pi inj/(2N+1)}.
\end{equation}
On a rather delicate aspect of path integrals for degrees of $\hat{\phi}$ and $\hat{\phi}^{\dagger}$, we will explain in the appendix in detail. By doing so, we are convinced that the model under consideration fulfills the Euclidicity postulate\cite{Abers-Lee,Coleman:85}, that is, the existence of the Euclidean path integral is the key to define a quantum system.

In order to check the validity of our formulation and the calculation above, let us consider the effective action defined by $Z^{(\mathrm{E})}[J]$ above. The classical solution for $\hat{q}_{0}$ under the effect of $J_{0}$ is determined by
\begin{equation}
	q^{\mathrm{cl.}}_{0;\,j}=\frac{1}{\epsilon}
	\frac{\partial W^{(\mathrm{E})}_{q_{0}}[J_{0}]}
	{\partial J_{0;\,j}}
	=-\epsilon\sum_{k=-N}^{N}\Delta^{(m)}_{j,\,k}J_{0;\,k}
\end{equation}
and can be inverted to yield
\begin{equation}
	J_{0;\,j}=-\frac{1}{\epsilon^{2}}
	\left\{-\tilde{\nabla}\nabla+(m\epsilon)^{2}\right\}
	q^{\mathrm{cl.}}_{0;\,j},
\end{equation}
where two kind of difference operator $\nabla f_{j}=f_{j}-f_{j-1}$ and $\tilde{\nabla}f_{j}=f_{j+1}-f_{j}$ have been introduced.
Substitution of the right hand side above into $W^{(\mathrm{E})}_{q_{0}}[J_{0}]$ results in
\begin{equation}
	W^{(\mathrm{E})}_{q_{0}}[J_{0}]=-\frac{1}{2\epsilon}\sum_{j=-N}^{N}
	q^{\mathrm{cl.}}_{0;\,j}\left\{-\tilde{\nabla}\nabla+(m\epsilon)^{2}\right\}
	q^{\mathrm{cl.}}_{0;\,j}.
\end{equation}
We also obtain
\begin{equation}
	\epsilon\sum_{j=-N}^{N}J_{0;\,j}q^{\mathrm{cl.}}_{0;\,j}=
	2W^{(\mathrm{E})}_{q_{0}}[J_{0}].
\end{equation}
We thus obtain from the $\hat{q}_{0}$ sector as the component of the effective action
\begin{equation}
	\varGamma^{(\mathrm{E})}_{q_{0}}[q^{\mathrm{cl.}}_{0}]=
	\frac{1}{2\epsilon}\sum_{j=-N}^{N}
	q^{\mathrm{cl.}}_{0;\,j}\left\{-\tilde{\nabla}\nabla+(m\epsilon)^{2}\right\}
	q^{\mathrm{cl.}}_{0;\,j}
\end{equation}
which is nothing but the Euclidean version of the classical action of this degree. In the similar manner, we obtain $\varGamma^{(\mathrm{E})}_{\phi}[\phi^{\mathrm{cl.}}]$ and $\varGamma^{(\mathrm{E})}_{\phi^{\dagger}}[\phi^{*\,\mathrm{cl.}}]$ as
\begin{equation}
	\varGamma^{(\mathrm{E})}_{\phi}[\phi^{\mathrm{cl.}}]=
	\frac{1}{2\epsilon}\sum_{j=-N}^{N}
	\phi^{\mathrm{cl.}}_{j}
	\left\{-\tilde{\nabla}\nabla+(\omega\epsilon)^{2}\right\}
	\phi^{\mathrm{cl.}}_{j}
\end{equation}
and
\begin{equation}
	\varGamma^{(\mathrm{E})}_{\phi^{\dagger}}[\phi^{\mathrm{*\,cl.}}]=
	\frac{1}{2\epsilon}\sum_{j=-N}^{N}
	\phi^{\mathrm{*\,cl.}}_{j}
	\left\{-\tilde{\nabla}\nabla+(\omega^{*}\epsilon)^{2}\right\}
	\phi^{\mathrm{*\,cl.}}_{j}.
\end{equation}
Thus we are convinced that, as far as the effective action is concerned, our formulation gives consistent result. If we aim to obtain the effective action for the original variable $q$, we arrange the source term to fit the form $J_{q}\hat{q}=J_{0}\hat{q}_{0}+J^{*}\hat{\phi}+J\hat{\phi}^{\dagger}$. Since operators in hand are redefined ones at the classical Lagrangian level, we have to carry out the inversion of their relations. To achieve this we put
\begin{equation}
	J_{0}=\frac{1}{\abs{W}}J_{q},\quad
	J^{*}=\frac{-i}{\sqrt{2W}}J_{q},\quad
	J=\frac{i}{\sqrt{2W^{*}}}J_{q}
\end{equation}
and rewrite the sum
\begin{equation}
	W^{(\mathrm{E})}[J_{q}]=
	W^{(\mathrm{E})}_{q_{0}}[J_{0}]+
	W^{(\mathrm{E})}_{\phi}[J^{*}]+
	W^{(\mathrm{E})}_{\phi^{\dagger}}[J]
\end{equation}
according to the change above. We then obtain, by writing $J_{q}$ simply as $J$,
\begin{equation}
	W^{(\mathrm{E})}[J]=-\frac{\epsilon^{2}}{2}J_{j}\Delta_{j,\,k}J_{k},
\end{equation}
where $\Delta_{j,\,k}$ being given by
\begin{equation}
\begin{aligned}
	\Delta_{j,\,k}=&\frac{1}{\abs{W}^{2}}\Delta^{(m)}_{j,\,k}-
	\frac{1}{2W}\Delta^{(\omega)}_{j,\,k}-
	\frac{1}{2W^{*}}\Delta^{(\omega^{*})}_{j,\,k}\\
	=&
	-\frac{1}{\epsilon^{2}}(M\epsilon)^{4}
	\left\{\Delta^{(m)}\Delta^{(\omega)}\Delta^{(\omega^{*})}\right\}_{j,\,k}.
\end{aligned}
\end{equation}
The classical solution $q^{\mathrm{cl.}}$ is determined by
\begin{equation}
	q^{\mathrm{cl.}}_{j}=\frac{1}{\epsilon}\frac{\partial W^{(\mathrm{E})}[J]}
	{\partial J_{j}}=-\epsilon\sum_{k=-N}^{N}\Delta_{j,\,k}J_{k}
\end{equation}
and whose inversion reads
\begin{equation}
	J_{j}=\frac{1}{\epsilon^{2}(M\epsilon)^{4}}
	\left\{-\tilde{\nabla}\nabla+(m\epsilon)^{2}\right\}
	\left\{-\tilde{\nabla}\nabla+(\omega\epsilon)^{2}\right\}
	\left\{-\tilde{\nabla}\nabla+(\omega^{*}\epsilon)^{2}\right\}
	q^{\mathrm{cl.}}_{j}.
\end{equation}
We can then express $W^{(\mathrm{E})}[J]$ in terms of $q^{\mathrm{cl.}}$ as
\begin{equation}
	W^{(\mathrm{E})}[J]=-\frac{1}{2\epsilon(M\epsilon)^{4}}
	\sum_{j=-N}^{N}q^{\mathrm{cl.}}_{j}
	\left\{-\tilde{\nabla}\nabla+(m\epsilon)^{2}\right\}
	\left\{-\tilde{\nabla}\nabla+(\omega\epsilon)^{2}\right\}
	\left\{-\tilde{\nabla}\nabla+(\omega^{*}\epsilon)^{2}\right\}
	q^{\mathrm{cl.}}_{j}.
\end{equation}
We also rewrite $\epsilon\sum_{j=-N}^{N}J_{j}q^{\mathrm{cl.}}_{j}$ to find
\begin{equation}
	\epsilon\sum_{j=-N}^{N}J_{j}q^{\mathrm{cl.}}_{j}=2W^{(\mathrm{E})}[J]
\end{equation}
then we obtain
\begin{equation}
\begin{aligned}
	\varGamma^{(\mathrm{E})}[q^{\mathrm{cl.}}]=&
	\frac{1}{2\epsilon(M\epsilon)^{4}}
	\sum_{j=-N}^{N}q^{\mathrm{cl.}}_{j}
	\left\{-\tilde{\nabla}\nabla+(m\epsilon)^{2}\right\}
	\left\{-\tilde{\nabla}\nabla+(\omega\epsilon)^{2}\right\}
	\left\{-\tilde{\nabla}\nabla+(\omega^{*}\epsilon)^{2}\right\}
	q^{\mathrm{cl.}}_{j}\\
	=&
	\frac{1}{2\epsilon}
	\sum_{j=-N}^{N}q^{\mathrm{cl.}}_{j}
	\left\{-\tilde{\nabla}\nabla+(m\epsilon)^{2}\right\}\left\{
	1+\frac{(\tilde{\nabla}\nabla)^{2}}{(M\epsilon)^{4}}\right\}
	q^{\mathrm{cl.}}_{j}
\end{aligned}
\end{equation}
which is again the discretized Euclidean version of the classical action of the system under consideration. We may therefore consider that our formulation is reliable at least for the tree level and that we may perform the perturbative expansion for a suitable potentiel $V(q)$ since we have the generating functional of Green's functions in hand.

Before closing this section, we should add a comment on the covariance of a path integral; as mentioned above, we can utilize the completness of coherent states as the basic tool for constructing path intergal representation. The same remains true for path integrals of field theories. If we formulate, however, path integrals in terms of coherent state, the action of such path integrals become naturally non-covariant. Nevertheless we can obtain covariant propagators or effective actions. It is therefore not the action of a path integral that determines the covariance of a path integral.

\section{Scalar Lee-Wick model}

To develop our method shown in the previous section to the quantum field theory, we here consider scalar Lee-Wick model defined by the Lagrangian density
\begin{equation}
	{\mathcal L}=-\frac{1}{2}\varphi\left(\DAlambert+m_{0}^{2}\right)
	\left(1+\frac{\DAlambert^{2}}{M_{0}^{4}}\right)\varphi-V(\varphi).
\end{equation}
Through the same proceder in the quantum mechanical model, we introduce
\begin{equation}
	\tilde{\phi}_{0}=\frac{1}{F{'}(s_{0})}
	\left(\DAlambert+iM_{0}^{2}\right)\left(\DAlambert-iM_{0}^{2}\right)\varphi,
\end{equation}
and
\begin{equation}
	\tilde{\phi}=\frac{1}{F{'}(s_{+})}
	\left(\DAlambert+m_{0}^{2}\right)\left(\DAlambert-iM_{0}^{2}\right)\varphi,
	\quad
	\tilde{\phi}^{\dagger}=\frac{1}{F{'}(s_{-})}
	\left(\DAlambert+m_{0}^{2}\right)\left(\DAlambert+iM_{0}^{2}\right)\varphi,
\end{equation}
where $F(s)$ is the same polynomial and $s_{0}=-m_{0}^{2}$ and $s_{\pm}=\mp iM_{0}^{2}$ as before. The Lagrangian will be then rewritten as
\begin{equation}
	{\mathcal L}=
	\frac{\abs{W}^{2}}{2}
	\left\{(\partial_{\mu}\tilde{\phi}_{0})^{2}-
	m_{0}^{2}\tilde{\phi}_{0}^{2}\right\}
	-W
	\left\{
	(\partial_{\mu}\tilde{\phi})^{2}-iM_{0}^{2}\tilde{\phi}^{2}\right\}
	-W^{*}\left\{
	(\partial_{\mu}\tilde{\phi})^{\dagger\,2}+
	iM_{0}^{2}\tilde{\phi}^{\dagger\,2}\right\}-V(\varphi),
\end{equation}
where $W$ is given by $1+im_{0}^{2}/M_{0}^{2}$ as before.
To simplify the notation, we redefine variables as
\begin{equation}
	\phi_{0}\equiv\abs{W}\tilde{\phi}_{0},\quad
	\phi\equiv i\sqrt{2W}\tilde{\phi},\quad
	\phi^{\dagger}\equiv-i\sqrt{2W^{*}}\tilde{\phi}^{\dagger}.
\end{equation}
We then obtain
\begin{equation}
	{\mathcal L}=\frac{1}{2}
	\left\{(\partial_{\mu}\phi_{0})^{2}-m_{0}^{2}\phi_{0}^{2}\right\}
	+\frac{1}{2}\left\{(\partial_{\mu}\phi)^{2}-iM_{0}^{2}\phi^{2}\right\}
	+\frac{1}{2}\left\{
	(\partial_{\mu}\phi^{\dagger})^{2}+iM_{0}^{2}\phi^{\dagger\,2}\right\}-
	V(\varphi).
\end{equation}
Thus the scalar field theory can be treated in the same way as we have studied quantum mechanical model. We here, however, introduce a spatial cubic lattice, whose length $L$ to be divided into $2N+1$ pieces of length $a$, to make field operators well-defined. Lattice points will be designated by a set of integers $n_{1},\,n_{2}$ and $n_{3}$ each taking values in $0,\,\pm1,\,\pm2,\,\dots,\,\pm N$ as $\bm{x}_{\bm{n}}=\bm{n}a$. We then consider the quantization of the free Lagrangian
\begin{equation}
\label{eq:lag0}
	L_{0}=\frac{1}{2a}\sum_{\bm{n}=-N}^{N}\left\{
	\nabla_{\mu}\bm{\varPhi}^{\mathrm{T}}_{\bm{n}}(x^{0})
	\nabla^{\mu}\bm{\varPhi}_{\bm{n}}(x^{0})-
	\bm{\varPhi}^{\mathrm{T}}_{\bm{n}}(x^{0})\bm{M}^{2}
	\bm{\varPhi}_{\bm{n}}(x^{0})\right\}
\end{equation}
where
\begin{equation}
	\bm{\varPhi}^{\mathrm{T}}_{\bm{n}}(x^{0})=\left(\phi_{0;\,\bm{n}}(x^{0}),\,
	\phi_{\bm{n}}(x^{0}),\,\phi^{\dagger}_{\bm{n}}(x^{0})\right)
\end{equation}
and dimensionless field operators are defined as $\phi_{\bm{n}}(x^{0})=a\phi(\bm{x}_{\bm{n}},x^{0})$ and the mass matrix $\bm{M}^{2}$ is a diagonal matrix whose components are given by dimensionless mass parameters, i.e. $m^{2}=m_{0}^{2}a^{2}$, $\omega^{2}=iM^{2}=iM_{0}^{2}a^{2}$ and $\omega^{*\,2}=-iM^{2}=-iM_{0}^{2}a^{2}$. The sum above is the abbreviation of sums with respect to $n_{k}=0,\,\pm1,\,\pm2,\,\dots,\,\pm N$ for $k=1,\,2,\,3$. The difference and differential operators in \eqref{eq:lag0} are defined as follows: for a field $f_{\bm{n}}(x^{0})$ on a lattice $\nabla_{k}$ ($k=1,\,2,\,3$) acts as $\nabla_{k}f_{\bm{n}}(x^{0})=f_{\bm{n}}(x^{0})-f_{\bm{n}-\hat{\bm{k}}}(x^{0})$, where $\hat{\bm{k}}$ is the unit vector in the $k$-th direction, while $\nabla_{0}$ is defined by $\nabla_{0}=a\partial_{0}$ here and for a while. When we formulate a time sliced path integral on the lattice, we will then switch the definition of $\nabla_{0}$ to the suitable one by changing the notation of the field to $f_{n}=f_{\bm{n}}(n_{0}a)$ or $f_{n}=f_{\bm{n}}(n_{4}a)$ corresponding to we are considering Minkowski space or Euclidean space. 

We assume the periodic boundary condition to field variables on the lattice and introduce orthogonal basis by $F_{\bm{n}}^{\bm{r}}=F_{n_{1}}^{r_{1}}F_{n_{2}}^{r_{2}}F_{n_{3}}^{r_{3}}$. The orthogonality and completeness of this basis read
\begin{equation}
	\sum_{\bm{n}=-N}^{N}F_{\bm{n}}^{\bm{r}\,*}F_{\bm{n}}^{\bm{r}{'}}
	=\delta_{\bm{r},\,\bm{r}{'}},\quad
	\sum_{\bm{r}=-N}^{N}F_{\bm{n}}^{\bm{r}}F_{\bm{n}{'}}^{\bm{r}\,*}
	=\delta_{\bm{n},\,\bm{n}{'}}.
\end{equation}

For the quantized system, the Hamiltonian is given by
\begin{equation}
\label{eq:ham0}
	\hat{H}_{0}=\frac{1}{2a}\sum_{\bm{n}=-N}^{N}\left\{
	\hat{\bm{\varPi}}^{\mathrm{T}}_{\bm{n}}\hat{\bm{\varPi}}_{\bm{n}}+
	\hat{\bm{\varPhi}}^{\mathrm{T}}_{\bm{n}}
	\left(-\Laplace+\bm{M}^{2}\right)
	\hat{\bm{\varPhi}}_{\bm{n}}\right\}
\end{equation}
where
\begin{equation}
	\hat{\bm{\varPi}}^{\mathrm{T}}_{\bm{n}}=\left(\hat{\varPi}_{0;\,\bm{n}},\,
	\hat{\varPi}_{\bm{n}},\,\hat{\varPi}^{\dagger}_{\bm{n}}\right)
\end{equation}
and the commutation relations are given by
\begin{equation}
	[\hat{\phi}_{0;\,\bm{n}},\,\hat{\varPi}_{0;\,\bm{n}{'}}]=
	[\hat{\phi}_{\bm{n}},\,\hat{\varPi}_{\bm{n}{'}}]=
	[\hat{\phi}_{\bm{n}}^{\dagger},\,\hat{\varPi}_{\bm{n}{'}}^{\dagger}]=
	i\delta_{\bm{n},\,\bm{n}{'}},
\end{equation}
in addition to the other trivial ones. Here all operators are Schr\"{o}dinger operators and we have written \eqref{eq:ham0} $\sum_{k=1}^{3}\tilde{\nabla}_{k}\nabla^{k}=-\Laplace$ for simplicity.

To construct Fock space representation of the canonical commutation relations, we expand field operators and their conjugates into the Fourier series as
\begin{equation}
\begin{gathered}
	\hat{\phi}_{0;\,\bm{n}}=\sum_{\bm{r}=-N}^{N}
	\frac{1}{\sqrt{2\omega_{0}(\bm{r})}}\left\{
	\hat{a}_{0}(\bm{r})F_{\bm{n}}^{\bm{r}}+
	\hat{a}_{0}^{\dagger}(\bm{r})F_{\bm{n}}^{\bm{r}\,*}\right\},\quad
	\hat{\varPi}_{0;\,\bm{n}}=-i\sum_{\bm{r}=-N}^{N}
	\sqrt{\frac{\omega_{0}(\bm{r})}{2}}\left\{
	\hat{a}_{0}(\bm{r})F_{\bm{n}}^{\bm{r}}-
	\hat{a}_{0}^{\dagger}(\bm{r})F_{\bm{n}}^{\bm{r}\,*}\right\},\\
	\hat{\phi}_{\bm{n}}=\sum_{\bm{r}=-N}^{N}
	\frac{1}{\sqrt{2\omega(\bm{r})}}\left\{
	\hat{\alpha}(\bm{r})F_{\bm{n}}^{\bm{r}}+
	\hat{\beta}^{\dagger}(\bm{r})F_{\bm{n}}^{\bm{r}\,*}\right\},\quad
	\hat{\varPi}_{\bm{n}}=-i\sum_{\bm{r}=-N}^{N}
	\sqrt{\frac{\omega(\bm{r})}{2}}\left\{
	\hat{\alpha}(\bm{r})F_{\bm{n}}^{\bm{r}}-
	\hat{\beta}^{\dagger}(\bm{r})F_{\bm{n}}^{\bm{r}\,*}\right\},\\
	\hat{\phi}^{\dagger}_{\bm{n}}=\sum_{\bm{r}=-N}^{N}
	\frac{1}{\sqrt{2\omega^{*}(\bm{r})}}\left\{
	\hat{\beta}(\bm{r})F_{\bm{n}}^{\bm{r}}+
	\hat{\alpha}^{\dagger}(\bm{r})F_{\bm{n}}^{\bm{r}\,*}\right\},\quad
	\hat{\varPi}^{\dagger}_{\bm{n}}=-i\sum_{\bm{r}=-N}^{N}
	\sqrt{\frac{\omega^{*}(\bm{r})}{2}}\left\{
	\hat{\beta}(\bm{r})F_{\bm{n}}^{\bm{r}}-
	\hat{\alpha}^{\dagger}(\bm{r})F_{\bm{n}}^{\bm{r}\,*}\right\},
\end{gathered}
\end{equation}
where
\begin{equation}
	\omega_{0}(\bm{r})=\frac{1}{a}\sqrt{
	\sum_{k=1}^{3}4\sin^{2}\left(\frac{\pi r_{k}}{2N+1}\right)+m^{2}}
\end{equation}
and
\begin{equation}
	\omega(\bm{r})=\frac{1}{a}\sqrt{
	\sum_{k=1}^{3}4\sin^{2}\left(\frac{\pi r_{k}}{2N+1}\right)+iM^{2}},\quad
	\omega^{*}(\bm{r})=\frac{1}{a}\sqrt{
	\sum_{k=1}^{3}4\sin^{2}\left(\frac{\pi r_{k}}{2N+1}\right)-iM^{2}}.
\end{equation}
Note that we should define square roots above so that real parts of both $\omega(\bm{r})$ and $\omega^{*}(\bm{r})$ to be positive. It is possible even for the zero mode; we define $\omega(0)=e^{i\pi/4}M$ and $\omega^{*}(0)=e^{-i\pi/4}M$.
By solving, for example
\begin{equation}
	\hat{\alpha}(\bm{r})=\sum_{\bm{n}=-N}^{N}F_{\bm{N}}^{\bm{r}\,*}
	\left\{\omega(\bm{r})\hat{\phi}_{\bm{n}}+i\hat{\varPi}_{\bm{n}}\right\},
\end{equation}
we find commutation relations of creation and annihilation operators given by
\begin{equation}
	[\hat{a}_{0}(\bm{r}),\,\hat{a}_{0}^{\dagger}(\bm{r}{'})]=
	[\hat{\alpha}(\bm{r}),\,\hat{\beta}^{\dagger}(\bm{r}{'})]=
	[\hat{\beta}(\bm{r}),\,\hat{\alpha}^{\dagger}(\bm{r}{'})]=
	\delta_{\bm{r},\,\bm{r}{'}}
\end{equation}
in addition to the trivial ones. The first one shows that we can quantize the degree of $\hat{\phi}_{0}$ on the positive definite Fock space. On the other hand, the second and the third ones indicates the need of indefinite metric representation.

To extend the method developed for quantum mechanical system in the previous section, we decompose field operators into the "positive frequency" parts and the "negative frequency" parts as
\begin{equation}
	\hat{\phi}_{0;\,\bm{n}}=\hat{\phi}_{0;\,\bm{n}}^{(+)}+
	\hat{\phi}_{0;\,\bm{n}}^{(-)},\quad
	\hat{\phi}_{\bm{n}}=\hat{\phi}_{\bm{n}}^{(+)}+
	\hat{\phi}_{\bm{n}}^{(-)},\quad
	\hat{\phi}_{\bm{n}}^{\dagger}=\hat{\phi}_{\bm{n}}^{\dagger\,(+)}+
	\hat{\phi}_{\bm{n}}^{\dagger\,(-)},
\end{equation}
where positive frequency parts designated by $(+)$ correspond to the series of $F_{\bm{n}}^{\bm{r}}$ while negative frequency parts are given by the series of $F_{\bm{n}}^{\bm{r}\,*}$. We write commutation relations of these operators as
\begin{equation}
	[\hat{\phi}_{0;\,\bm{n}}^{(+)},\hat{\phi}_{0;\,\bm{n}{'}}^{(-)}]=
	\frac{1}{2}K^{-1}_{(\omega_{0});\,\bm{n},\,\bm{n}{'}},\quad
	[\hat{\phi}_{\bm{n}}^{(+)},\hat{\phi}_{\bm{n}{'}}^{(-)}]=
	\frac{1}{2}K^{-1}_{(\omega);\,\bm{n},\,\bm{n}{'}},\quad
	[\hat{\phi}_{\bm{n}}^{\dagger\,(+)},\hat{\phi}_{\bm{n}{'}}^{\dagger\,(-)}]=
	\frac{1}{2}K^{-1}_{(\omega^{*});\,\bm{n},\,\bm{n}{'}},
\end{equation}
where
\begin{equation}
	K^{-1}_{(\omega_{0});\,\bm{n},\,\bm{n}{'}}=\sum_{\bm{r}=-N}^{N}
	\frac{1}{\omega_{0}(\bm{r})}
	F_{\bm{n}}^{\bm{r}}F_{\bm{n}{'}}^{\bm{r}\,*},\quad
	K^{-1}_{(\omega);\,\bm{n},\,\bm{n}{'}}=\sum_{\bm{r}=-N}^{N}
	\frac{1}{\omega(\bm{r})}F_{\bm{n}}^{\bm{r}}F_{\bm{n}{'}}^{\bm{r}\,*},\quad
	K^{-1}_{(\omega^{*});\,\bm{n},\,\bm{n}{'}}=\sum_{\bm{r}=-N}^{N}
	\frac{1}{\omega^{*}(\bm{r})}F_{\bm{n}}^{\bm{r}}F_{\bm{n}{'}}^{\bm{r}\,*}.
\end{equation}

On defining the vacuum $\vert0\rangle$ to obey
\begin{equation}
	\hat{\phi}_{0;\,\bm{n}}^{(+)}\vert0\rangle=
	\hat{\phi}_{\bm{n}}^{(+)}\vert0\rangle=
	\hat{\phi}_{\bm{n}}^{\dagger\,(+)}\vert0\rangle=0,\quad
	\langle0\vert0\rangle=1,
\end{equation}
we introduce simultaneous eigenvectors (coherent states) of $\hat{\phi}_{0;\,\bm{n}}^{(+)}$, $\hat{\phi}_{\bm{n}}^{(+)}$ and $\hat{\phi}_{\bm{n}}^{\dagger\,(+)}$ by
\begin{equation}
	\vert\{\varPhi^{(+)}\}\rangle=
	\exp\left[2\sum_{\bm{n},\,\bm{n}{'}=-N}^{N}\left\{
	\hat{\phi}_{0;\,\bm{n}}^{(-)}K^{(\omega_{0})}_{\bm{n},\,\bm{n}{'}}
	\phi_{0;\,\bm{n}{'}}^{(+)}+
	\hat{\phi}_{\bm{n}}^{(-)}K^{(\omega)}_{\bm{n},\,\bm{n}{'}}
	\phi_{\bm{n}{'}}^{(+)}+
	\hat{\phi}_{\bm{n}}^{\dagger\,(-)}K^{(\omega^{*})}_{\bm{n},\,\bm{n}{'}}
	\phi_{\bm{n}{'}}^{\dagger\,(+)}\right\}\right]
	\vert0\rangle,
\end{equation}
where $K^{(\omega_{0})}$, $K^{(\omega)}$ and $K^{(\omega^{*})}$ are inverse of $K^{-1}_{(\omega_{0})}$, $K^{-1}_{(\omega)}$ and $K^{-1}_{(\omega^{*})}$, respectively, and $c$-number fields $\phi_{0;\,\bm{n}}^{(+)}$, $\phi_{\bm{n}}^{(+)}$ and $\phi_{\bm{n}}^{\dagger\,(+)}$ give eigenvalues of $\hat{\phi}_{0;\,\bm{n}}^{(+)}$, $\hat{\phi}_{\bm{n}}^{(+)}$ and $\hat{\phi}_{\bm{n}}^{\dagger\,(+)}$, respectively, in this order.
In the same way, we define its dual by
\begin{equation}
	\langle\underline{\{\varPhi^{(-)}\}}\vert=\langle0\vert
	\exp\left[2\sum_{\bm{n},\,\bm{n}{'}=-N}^{N}\left\{
	\phi_{0;\,\bm{n}}^{(-)}K^{(\omega_{0})}_{\bm{n},\,\bm{n}{'}}
	\hat{\phi}_{0;\,\bm{n}{'}}^{(+)}+
	\phi_{\bm{n}}^{(-)}K^{(\omega)}_{\bm{n},\,\bm{n}{'}}
	\hat{\phi}_{\bm{n}{'}}^{(+)}+
	\phi_{\bm{n}}^{\dagger\,(-)}K^{(\omega^{*})}_{\bm{n},\,\bm{n}{'}}
	\hat{\phi}_{\bm{n}{'}}^{\dagger\,(+)}\right\}\right]
\end{equation}
upon which $\hat{\phi}_{0;\,\bm{n}}^{(-)}$, $\hat{\phi}_{\bm{n}}^{(-)}$ and $\hat{\phi}_{\bm{n}}^{\dagger\,(-)}$ take eigenvalues $\phi_{0;\,\bm{n}}^{(-)}$, $\phi_{\bm{n}}^{(-)}$ and $\phi_{\bm{n}}^{\dagger\,(-)}$, respectively.
Note that $\langle\underline{\{\varPhi^{(-)}\}}\vert\ne\vert\{\varPhi^{(+)}\}\rangle^{\dagger}$.
The inner product between the states above can be easily obtained as
\begin{equation}
	\langle\underline{\{\varPhi^{(-)}\}}\vert\{\varPhi{'}^{(+)}\}\rangle=
	\exp\left[2\sum_{\bm{n},\,\bm{n}{'}=-N}^{N}\left\{
	\phi_{0;\,\bm{n}}^{(-)}K^{(\omega_{0})}_{\bm{n},\,\bm{n}{'}}
	\phi{'}_{0;\,\bm{n}{'}}^{(+)}+
	\phi_{\bm{n}}^{(-)}K^{(\omega)}_{\bm{n},\,\bm{n}{'}}
	\phi{'}_{\bm{n}{'}}^{(+)}+
	\phi_{\bm{n}}^{\dagger\,(-)}K^{(\omega^{*})}_{\bm{n},\,\bm{n}{'}}
	\phi{'}_{\bm{n}{'}}^{\dagger\,(+)}\right\}\right].
\end{equation}
The proof of the completness of the coherent state above is straightforward but we omit it here because we do not make use of it in this paper.

We now consider
\begin{multline}
	\hat{\varPi}[\{\varPhi\}]=
	\frac{1}{(2\pi)^{3(2N+1)^{3}}}\int\!\!\prod_{\bm{n}=-N}^{N}
	d\lambda_{0;\,\bm{n}}\,d\lambda_{\bm{n}}\,d\lambda_{\bm{n}}^{*}\,\\
	\times
	\exp\left[i\sum_{\bm{n}=-N}^{N}\left\{
	\lambda_{0;\,\bm{n}}\left(\hat{\phi}_{0;\,\bm{n}}-\phi_{0;\,\bm{n}}\right)+
	\lambda_{\bm{n}}^{*}\left(\hat{\phi}_{\bm{n}}-\phi_{\bm{n}}\right)+
	\lambda_{\bm{n}}\left(\hat{\phi}_{\bm{n}}^{\dagger}-
	\phi_{\bm{n}}^{\dagger}\right)\right\}\right]
\end{multline}
and rewrite it as
\begin{multline}
	\hat{\varPi}[\{\varPhi\}]=
	\frac{1}{(2\pi)^{3(2N+1)^{3}}}\int\!\!\prod_{\bm{n}=-N}^{N}
	d\lambda_{0;\,\bm{n}}\,d\lambda_{\bm{n}}\,d\lambda_{\bm{n}}^{*}\,
	\exp\left\{-
	i\sum_{\bm{n}=-N}^{N}\left(
	\lambda_{0;\,\bm{n}}\phi_{0;\,\bm{n}}+
	\lambda_{\bm{n}}^{*}\phi_{\bm{n}}+
	\lambda_{\bm{n}}\phi_{\bm{n}}^{\dagger}\right)\right\}\\
	\times
	\exp\left\{-\frac{1}{4}\sum_{\bm{n},\,\bm{n}{'}=-N}^{N}\left(
	\lambda_{0;\,\bm{n}}K^{-1}_{(\omega_{0});\,\bm{n},\,\bm{n}{'}}
	\lambda_{0;\,\bm{n}{'}}+
	\lambda_{\bm{n}}^{*}K^{-1}_{(\omega);\,\bm{n},\,\bm{n}{'}}
	\lambda_{\bm{n}{'}}^{*}+
	\lambda_{\bm{n}}K^{-1}_{(\omega^{*});\,\bm{n},\,\bm{n}{'}}
	\lambda_{\bm{n}{'}}\right)
	\right\}\\
	\times
	\exp\left\{i\sum_{\bm{n}=-N}^{N}\left(
	\lambda_{0;\,\bm{n}}\hat{\phi}_{0;\,\bm{n}}^{(-)}+
	\lambda_{\bm{n}}^{*}\hat{\phi}_{\bm{n}}^{(-)}+
	\lambda_{\bm{n}}\hat{\phi}_{\bm{n}}^{\dagger\,(-)}\right)\right\}
	\exp\left\{i\sum_{\bm{n}=-N}^{N}\left(
	\lambda_{0;\,\bm{n}}\hat{\phi}_{0;\,\bm{n}}^{(+)}+
	\lambda_{\bm{n}}^{*}\hat{\phi}_{\bm{n}}^{(+)}+
	\lambda_{\bm{n}}\hat{\phi}_{\bm{n}}^{\dagger\,(+)}\right)\right\}.
\end{multline}
We then sandwich $\hat{\varPi}[\{\varPhi\}]$ between coherent states defined above to obtain
\begin{multline}
	\langle\underline{\{\varPhi^{(-)}\}}\vert\hat{\varPi}[\{\varPhi\}]
	\vert\{\varPhi{'}^{(+)}\}\rangle
	=
	\frac{1}{(2\pi)^{3(2N+1)^{3}}}\int\!\!\prod_{\bm{n}=-N}^{N}
	d\lambda_{0;\,\bm{n}}\,d\lambda_{\bm{n}}\,d\lambda_{\bm{n}}^{*}\\
	\times
	\exp\left\{-\frac{1}{4}\sum_{\bm{n},\,\bm{n}{'}=-N}^{N}\left(
	\lambda_{0;\,\bm{n}}K^{-1}_{(\omega_{0});\,\bm{n},\,\bm{n}{'}}
	\lambda_{0;\,\bm{n}{'}}+
	\lambda_{\bm{n}}^{*}K^{-1}_{(\omega);\,\bm{n},\,\bm{n}{'}}
	\lambda_{\bm{n}{'}}^{*}+
	\lambda_{\bm{n}}K^{-1}_{(\omega^{*});\,\bm{n},\,\bm{n}{'}}
	\lambda_{\bm{n}{'}}\right)
	\right\}\\
	\times
	\exp\left[i\sum_{\bm{n}=-N}^{N}\left\{
	\lambda_{0;\,\bm{n}}
	\left(
	\phi_{0;\,\bm{n}}^{(+)}+\phi{'}_{0;\,\bm{n}}^{(-)}-\phi_{0;\,\bm{n}}\right)+
	\lambda_{\bm{n}}^{*}\left(
	\phi_{\bm{n}}^{(+)}+\phi{'}_{\bm{n}}^{(-)}-\phi_{\bm{n}}\right)+
	\lambda_{\bm{n}}\left(
	\phi_{\bm{n}}^{\dagger\,(+)}+\phi{'}_{\bm{n}}^{\dagger\,(-)}-
	\phi_{\bm{n}}^{\dagger}\right)\right\}\right]\\
	\times
	\langle\{\varPhi^{(-)}\}\vert\{\varPhi{'}^{(+)}\}\rangle.
\end{multline}
By carrying out the Gaussian integrations, we find
\begin{multline}
\label{eq:res01}
	\langle\underline{\{\varPhi^{(-)}\}}\vert\hat{\varPi}[\{\varPhi\}]
	\vert\{\varPhi{'}^{(+)}\}\rangle=
	\frac{1}{\sqrt{\det(\pi^{3}
	K^{-1}_{(\omega_{0})}K^{-1}_{(\omega)}K^{-1}_{(\omega^{*})})}}\\
	\times
	\exp\left\{-\sum_{\bm{n},\,\bm{n}{'}=-N}^{N}\left(
	X_{0;\,\bm{n}}K^{(\omega_{0})}_{\bm{n},\,\bm{n}{'}}X_{0;\,\bm{n}{'}}+
	X_{\bm{n}}K^{(\omega)}_{\bm{n},\,\bm{n}{'}}X_{\bm{n}{'}}+
	X_{\bm{n}}^{\dagger}K^{(\omega^{*})}_{\bm{n},\,\bm{n}{'}}
	X_{\bm{n}{'}}^{\dagger}
	\right)\right\}
	\langle\{\varPhi^{(-)}\}\vert\{\varPhi{'}^{(+)}\}\rangle,
\end{multline}
where
\begin{equation}
	X_{0;\,\bm{n}}=
	\phi_{0;\,\bm{n}}^{(+)}+\phi{'}_{0;\,\bm{n}}^{(-)}-\phi_{0;\,\bm{n}},\quad
	X_{\bm{n}}=\phi_{\bm{n}}^{(+)}+\phi{'}_{\bm{n}}^{(-)}-\phi_{\bm{n}},\quad
	X_{\bm{n}}^{\dagger}=
	\phi_{\bm{n}}^{\dagger\,(+)}+\phi{'}_{\bm{n}}^{\dagger\,(-)}-
	\phi_{\bm{n}}^{\dagger}.
\end{equation}
Expanding products in the exponent and remembering the inner product between coherent states, we find
\begin{equation}
	\langle\{\varPhi^{(-)}\}\vert\hat{\varPi}[\{\varPhi\}]
	\vert\{\varPhi{'}^{(+)}\}\rangle=
	\langle\underline{\{\varPhi^{(-)}\}}\vert\{\varPhi\}\rangle
	\langle\underline{\{\varPhi\}}\vert\{\varPhi{'}^{(+)}\}\rangle
\end{equation}
where simultaneous eigenvectors of $\hat{\phi}_{0;\,\bm{n}}$, $\hat{\phi}_{\bm{n}}$ and $\hat{\phi}^{\dagger}_{\bm{n}}$ are given by
\begin{multline}
	\vert\{\varPhi\}\rangle=
	\frac{1}{\{\det(\pi^{3}
	K^{-1}_{(\omega_{0})}K^{-1}_{(\omega)}K^{-1}_{(\omega^{*})})\}^{1/4}}
	\exp\left\{-\frac{1}{2}\sum_{\bm{n},\,\bm{n}{'}=-N}^{N}\left(
	\phi_{0;\,\bm{n}}K^{(\omega_{0})}_{\bm{n},\,\bm{n}{'}}\phi_{0;\,\bm{n}{'}}+
	\phi_{\bm{n}}K^{(\omega)}_{\bm{n},\,\bm{n}{'}}\phi_{\bm{n}{'}}+
	\phi_{\bm{n}}^{\dagger}K^{(\omega^{*})}_{\bm{n},\,\bm{n}{'}}
	\phi_{\bm{n}{'}}^{\dagger}\right)\right\}\\
	\times
	\exp\left\{2\sum_{\bm{n},\,\bm{n}{'}=-N}^{N}\left(
	\hat{\phi}_{0;\,\bm{n}}^{(-)}K^{(\omega_{0})}_{\bm{n},\,\bm{n}{'}}
	\phi_{0;\,\bm{n}{'}}+
	\hat{\phi}_{\bm{n}}^{(-)}K^{(\omega)}_{\bm{n},\,\bm{n}{'}}\phi_{\bm{n}{'}}+
	\hat{\phi}_{\bm{n}}^{\dagger\,(-)}K^{(\omega^{*})}_{\bm{n},\,\bm{n}{'}}
	\phi_{\bm{n{'}}}^{\dagger}\right)\right\}\\
	\exp\left\{-\sum_{\bm{n},\,\bm{n}{'}=-N}^{N}\left(
	\hat{\phi}_{0;\,\bm{n}}^{(-)}K^{(\omega_{0})}_{\bm{n},\,\bm{n}{'}}
	\hat{\phi}_{0;\,\bm{n}{'}}^{(-)}+
	\hat{\phi}_{\bm{n}}^{(-)}K^{(\omega)}_{\bm{n},\,\bm{n}{'}}
	\hat{\phi}_{\bm{n}{'}}^{(-)}+
	\hat{\phi}_{\bm{n}}^{\dagger\,(-)}K^{(\omega^{*})}_{\bm{n},\,\bm{n}{'}}
	\hat{\phi}_{\bm{n}{'}}^{\dagger\,(-)}\right)\right\}
	\vert0\rangle
\end{multline}
and by
\begin{multline}
	\langle\underline{\{\varPhi\}}\vert=
	\frac{1}{\{\det(\pi^{3}
	K^{-1}_{(\omega_{0})}K^{-1}_{(\omega)}K^{-1}_{(\omega^{*})})\}^{1/4}}
	\exp\left\{-\frac{1}{2}\sum_{\bm{n},\,\bm{n}{'}=-N}^{N}\left(
	\phi_{0;\,\bm{n}}K^{(\omega_{0})}_{\bm{n},\,\bm{n}{'}}\phi_{0;\,\bm{n}{'}}+
	\phi_{\bm{n}}K^{(\omega)}_{\bm{n},\,\bm{n}{'}}\phi_{\bm{n}{'}}+
	\phi_{\bm{n}}^{\dagger}K^{(\omega^{*})}_{\bm{n},\,\bm{n}{'}}
	\phi_{\bm{n}{'}}^{\dagger}\right)\right\}\\
	\times
	\langle0\vert
	\exp\left\{2\sum_{\bm{n},\,\bm{n}{'}=-N}^{N}\left(
	\phi_{0;\,\bm{n}}K^{(\omega_{0})}_{\bm{n},\,\bm{n}{'}}
	\hat{\phi}_{0;\,\bm{n}{'}}^{(+)}+
	\phi_{\bm{n}}K^{(\omega)}_{\bm{n},\,\bm{n}{'}}\hat{\phi}_{\bm{n}{'}}^{(+)}+
	\phi_{\bm{n}}^{\dagger}K^{(\omega^{*})}_{\bm{n},\,\bm{n}{'}}
	\hat{\phi}_{\bm{n}{'}}^{\dagger\,(+)}\right)\right\}\\
	\exp\left\{-\sum_{\bm{n},\,\bm{n}{'}=-N}^{N}\left(
	\hat{\phi}_{0;\,\bm{n}}^{(+)}K^{(\omega_{0})}_{\bm{n},\,\bm{n}{'}}
	\hat{\phi}_{0;\,\bm{n}{'}}^{(+)}+
	\hat{\phi}_{\bm{n}}^{(+)}K^{(\omega)}_{\bm{n},\,\bm{n}{'}}
	\hat{\phi}_{\bm{n}{'}}^{(+)}+
	\hat{\phi}_{\bm{n}}^{\dagger\,(+)}K^{(\omega^{*})}_{\bm{n},\,\bm{n}{'}}
	\hat{\phi}_{\bm{n}{'}}^{\dagger\,(+)}\right)\right\}.
\end{multline}
Since we can integrate the right hand side of \eqref{eq:res01}, we observe
\begin{equation}
	\int\!\!\prod_{\bm{n}=-N}^{N}d\phi_{0;\,\bm{n}}\,d\phi_{\bm{n}}\,
	d\phi^{\dagger}_{\bm{n}}\,\hat{\varPi}[\{\varPhi\}]=
	\int\!\!\prod_{\bm{n}=-N}^{N}d\phi_{0;\,\bm{n}}\,d\phi_{\bm{n}}\,
	d\phi^{\dagger}_{\bm{n}}\,\vert\{\varPhi\}\rangle
	\langle\underline{\{\varPhi\}}\vert=1,
\end{equation}
where the integration contour can be taken along the real axis for all variables. Let us here calculate $\langle\underline{\{\varPhi\}}\vert\{\varPhi{'}\}\rangle$. To achieve this, we rewrite $\vert\{\varPhi{'}\}\rangle$ by making use of Gaussian identity as
\begin{multline}
	\vert\{\varPhi\}\rangle=\frac{1}{\{\det(\pi^{3}
	K^{-1}_{(\omega_{0})}K^{-1}_{(\omega)}K^{-1}_{(\omega^{*})})\}^{1/4}%
	\{\det(\pi^{3}K^{(\omega_{0})}K^{(\omega)}K^{(\omega^{*})})\}^{1/2}}
	\int\!\!\prod_{\bm{n}=-N}^{N}d\lambda_{0;\,\bm{n}}\,d\lambda_{\bm{n}}^{*}\,
	d\lambda_{\bm{n}}\\
	\times
	\exp\left\{\frac{1}{2}\sum_{\bm{n},\,\bm{n}{'}=-N}^{N}\left(
	\phi{'}_{0;\,\bm{n}}K^{(\omega_{0})}_{\bm{n},\,\bm{n}{'}}
	\phi{'}_{0;\,\bm{n}{'}}+
	\phi{'}_{\bm{n}}K^{(\omega)}_{\bm{n},\,\bm{n}{'}}\phi{'}_{\bm{n}{'}}+
	\phi{'}_{\bm{n}}^{\dagger}K^{(\omega^{*})}_{\bm{n},\,\bm{n}{'}}
	\phi{'}_{\bm{n}{'}}^{\dagger}\right)\right\}\\
	\times
	\exp\left\{-\sum_{\bm{n},\,\bm{n}{'}=-N}^{N}\left(
	\lambda_{0;\,\bm{n}}K^{-1}_{(\omega_{0});\,\bm{n},\,\bm{n}{'}}
	\lambda_{0;\,\bm{n}{'}}+
	\lambda_{\bm{n}}^{*}K^{-1}_{(\omega);\,\bm{n},\,\bm{n}{'}}
	\lambda_{\bm{n}{'}}^{*}+
	\lambda_{\bm{n}}K^{-1}_{(\omega^{*});\,\bm{n},\,\bm{n}{'}}
	\lambda_{\bm{n}{'}}\right)
	\right\}\\
	\exp\left[2i\sum_{\bm{n}=-N}^{N}\left\{
	\lambda_{0;\,\bm{n}}
	\left(
	\hat{\phi}_{0;\,\bm{n}}^{(-)}-\phi{'}_{0;\,\bm{n}}\right)+
	\lambda_{\bm{n}}^{*}\left(
	\hat{\phi}_{\bm{n}}^{(-)}-\phi{'}_{\bm{n}}\right)+
	\lambda_{\bm{n}}\left(
	\hat{\phi}_{\bm{n}}^{\dagger\,(-)}-
	\phi{'}_{\bm{n}}^{\dagger}\right)\right\}\right]
	\vert0\rangle,
\end{multline}
in which we can interpret
\begin{equation}
	\exp\left\{2i\sum_{\bm{n}=-N}^{N}\left(
	\lambda_{0;\,\bm{n}}\hat{\phi}_{0;\,\bm{n}}^{(-)}+
	\lambda_{\bm{n}}^{*}\hat{\phi}_{\bm{n}}^{(-)}+
	\lambda_{\bm{n}}\hat{\phi}_{\bm{n}}^{\dagger\,(-)}\right)\right\}
	\vert0\rangle=
	\vert\{\varPhi^{(+)}[\lambda_{0},\,\lambda^{*},\,\lambda]\}\rangle
\end{equation}
by identifying the eigenvalues of $\hat{\phi}_{0;\,\bm{n}}^{(+)}$, $\hat{\phi}_{\bm{n}}^{(+)}$ and $\hat{\phi}^{\dagger\,(+)}_{\bm{n}}$ on $\vert\{\varPhi^{(+)}[\lambda_{0},\,\lambda^{*},\,\lambda]\}\rangle$ as
\begin{equation}
	\phi_{0;\bm{n}}^{(+)}=i\sum_{\bm{n}{'}=-N}^{N}
	K^{-1}_{(\omega_{0});\,\bm{n},\,\bm{n}{'}}
	\lambda_{0;\,\bm{n}{'}},\quad
	\phi_{\bm{n}}^{(+)}=i\sum_{\bm{n}{'}=-N}^{N}
	K^{-1}_{(\omega);\,\bm{n},\,\bm{n}{'}}
	\lambda_{\bm{n}{'}}^{*},\quad
	\phi_{\bm{n}}^{\dagger\,(+)}=i\sum_{\bm{n}{'}=-N}^{N}
	K^{-1}_{(\omega^{*});\,\bm{n},\,\bm{n}{'}}
	\lambda_{\bm{n}{'}}.
\end{equation}
We then immediately obtain
\begin{equation}
\begin{aligned}
	\langle\underline{\{\varPhi\}}\vert\{\varPhi{'}\}\rangle=&
	\frac{1}{(2\pi)^{3(2N+1)^{3}}}\int\!\!\prod_{\bm{n}=-N}^{N}
	d\lambda_{0;\,\bm{n}}\,d\lambda_{\bm{n}}^{*}\,d\lambda_{\bm{n}}\\
	&\times
	\exp\left[i\sum_{\bm{n}=-N}^{N}\left\{
	\lambda_{0;\,\bm{n}}
	\left(
	\phi_{0;\,\bm{n}}-\phi{'}_{0;\,\bm{n}}\right)+
	\lambda_{\bm{n}}^{*}\left(
	\phi_{\bm{n}}-\phi{'}_{\bm{n}}\right)+
	\lambda_{\bm{n}}\left(
	\phi_{\bm{n}}^{\dagger}-
	\phi{'}_{\bm{n}}^{\dagger}\right)\right\}\right]\\
	=&
	\prod_{\bm{n}=-N}^{N}\delta\left(
	\phi_{0;\,\bm{n}}-\phi{'}_{0;\,\bm{n}}\right)
	\delta\left(
	\phi_{\bm{n}}-\phi{'}_{\bm{n}}\right)
	\delta\left(
	\phi_{\bm{n}}^{\dagger}-
	\phi{'}_{\bm{n}}^{\dagger}\right)
\end{aligned}
\end{equation}
in which the second and the third delta functions are complex delta functions in general.

Finding eigenvectors of $\hat{\varPi}_{0;\,\bm{n}}$, $\hat{\varPi}_{\bm{n}}$ and $\hat{\varPi}_{\bm{n}}^{\dagger}$ can be done in a same way but there is a quick way of finding it by assuming that the inner product between eigenvectors of field operators and their canonical conjugates must be the plane wave:
\begin{equation}
	\langle\underline{\{\varPhi\}}\vert\{\varPi\}\rangle=
	\frac{1}{\sqrt{(2\pi)^{3(2N+1)^{3}}}}
	\exp\left\{i\sum_{\bm{n}=-N}^{N}\left(\phi_{0;\,\bm{n}}\varPi_{0;\,\bm{n}}+
	\phi_{\bm{n}}\varPi_{\bm{n}}+
	\phi_{\bm{n}}^{\dagger}\varPi_{\bm{n}}^{\dagger}\right)\right\}.
\end{equation}
We then calculate the Fourier transform
\begin{equation}
	\vert\{\varPi\}\rangle=\int\!\!\prod_{\bm{n}=-N}^{N}
	d\phi_{0;\,\bm{n}}\,d\phi_{\bm{n}}\,d\phi_{\bm{n}}^{\dagger}\,
	\vert\{\varPhi\}\rangle\langle\underline{\{\varPhi\}}\vert\{\varPi\}\rangle
\end{equation}
to obtain
\begin{equation}
\begin{aligned}
	\vert\{\varPi\}\rangle=&
	\frac{1}{\{\det(\pi^{3}
	K^{(\omega_{0})}K^{(\omega)}K^{(\omega^{*})})\}^{1/4}}\\
	&\times
	\exp\left\{-\frac{1}{2}\sum_{\bm{n},\,\bm{n}{'}=-N}^{N}\left(
	\varPi_{0;\,\bm{n}}K^{-1}_{(\omega_{0});\,\bm{n},\,\bm{n}{'}}
	\varPi_{0;\,\bm{n}{'}}+
	\varPi_{\bm{n}}K^{-1}_{(\omega);\,\bm{n},\,\bm{n}{'}}\varPi_{\bm{n}{'}}+
	\varPi_{\bm{n}}^{\dagger}K^{-1}_{(\omega^{*});\,\bm{n},\,\bm{n}{'}}
	\varPi_{\bm{n}{'}}^{\dagger}\right)\right\}\\
	&\times
	\exp\left\{2\sum_{\bm{n},\,\bm{n}{'}=-N}^{N}\left(
	\hat{\varPi}_{0;\,\bm{n}}^{(-)}K^{-1}_{(\omega_{0});\,\bm{n},\,\bm{n}{'}}
	\varPi_{0;\,\bm{n}{'}}+
	\hat{\varPi}_{\bm{n}}^{(-)}K^{-1}_{(\omega);\,\bm{n},\,\bm{n}{'}}
	\varPi_{\bm{n}{'}}+
	\hat{\varPi}_{\bm{n}}^{\dagger\,(-)}
	K^{-1}_{(\omega^{*});\,\bm{n},\,\bm{n}{'}}
	\varPi_{\bm{n{'}}}^{\dagger}\right)\right\}\\
	&\times
	\exp\left\{-\sum_{\bm{n},\,\bm{n}{'}=-N}^{N}\left(
	\hat{\varPi}_{0;\,\bm{n}}^{(-)}K^{-1}_{(\omega_{0});\,\bm{n},\,\bm{n}{'}}
	\hat{\varPi}_{0;\,\bm{n}{'}}^{(-)}+
	\hat{\varPi}_{\bm{n}}^{(-)}K^{-1}_{(\omega);\,\bm{n},\,\bm{n}{'}}
	\hat{\varPi}_{\bm{n}{'}}^{(-)}+
	\hat{\varPi}_{\bm{n}}^{\dagger\,(-)}
	K^{-1}_{(\omega^{*});\,\bm{n},\,\bm{n}{'}}
	\hat{\varPi}_{\bm{n}{'}}^{\dagger\,(-)}\right)\right\}
	\vert0\rangle
\end{aligned}
\end{equation}
and also whose conjugate as
\begin{equation}
\begin{aligned}
	\langle\underline{\{\varPi\}}\vert=&
	\frac{1}{\{\det(\pi^{3}
	K^{(\omega_{0})}K^{(\omega)}K^{(\omega^{*})})\}^{1/4}}\\
	&\times
	\exp\left\{-\frac{1}{2}\sum_{\bm{n},\,\bm{n}{'}=-N}^{N}\left(
	\varPi_{0;\,\bm{n}}K^{-1}_{(\omega_{0});\,\bm{n},\,\bm{n}{'}}
	\varPi_{0;\,\bm{n}{'}}+
	\varPi_{\bm{n}}K^{-1}_{(\omega);\,\bm{n},\,\bm{n}{'}}\varPi_{\bm{n}{'}}+
	\varPi_{\bm{n}}^{\dagger}K^{-1}_{(\omega^{*});\,\bm{n},\,\bm{n}{'}}
	\varPi_{\bm{n}{'}}^{\dagger}\right)\right\}\\
	&\times
	\langle0\vert
	\exp\left\{2\sum_{\bm{n},\,\bm{n}{'}=-N}^{N}\left(
	\varPi_{0;\,\bm{n}}K^{-1}_{(\omega_{0});\,\bm{n},\,\bm{n}{'}}
	\hat{\varPi}_{0;\,\bm{n}{'}}^{(+)}+
	\varPi_{\bm{n}}K^{-1}_{(\omega);\,\bm{n},\,\bm{n}{'}}
	\hat{\varPi}_{\bm{n}{'}}^{(+)}+
	\varPi_{\bm{n}}^{\dagger}K^{-1}_{(\omega^{*});\,\bm{n},\,\bm{n}{'}}
	\hat{\varPi}_{\bm{n}{'}}^{\dagger\,(+)}\right)\right\}\\
	&\times
	\exp\left\{-\sum_{\bm{n},\,\bm{n}{'}=-N}^{N}\left(
	\hat{\varPi}_{0;\,\bm{n}}^{(+)}K^{-1}_{(\omega_{0});\,\bm{n},\,\bm{n}{'}}
	\hat{\varPi}_{0;\,\bm{n}{'}}^{(+)}+
	\hat{\varPi}_{\bm{n}}^{(+)}K^{-1}_{(\omega);\,\bm{n},\,\bm{n}{'}}
	\hat{\varPi}_{\bm{n}{'}}^{(+)}+
	\hat{\varPi}_{\bm{n}}^{\dagger\,(+)}
	K^{-1}_{(\omega^{*});\,\bm{n},\,\bm{n}{'}}
	\hat{\varPi}_{\bm{n}{'}}^{\dagger\,(+)}\right)\right\}.
\end{aligned}
\end{equation}
Here use has been made of
\begin{equation}
	\hat{\phi}_{0;\,\bm{n}}^{(-)}=i\sum_{\bm{n}=-N}^{N}
	K^{-1}_{(\omega_{0});\,\bm{n},\,\bm{n}{'}}
	\hat{\varPi}_{0;\,\bm{n}{'}}^{(-)},\quad
	\hat{\phi}_{\bm{n}}^{(-)}=i\sum_{\bm{n}=-N}^{N}
	K^{-1}_{(\omega);\,\bm{n},\,\bm{n}{'}}
	\hat{\varPi}_{\bm{n}{'}}^{(-)},\quad
	\hat{\phi}_{\bm{n}}^{\dagger\,(-)}=i\sum_{\bm{n}=-N}^{N}
	K^{-1}_{(\omega^{*});\,\bm{n},\,\bm{n}{'}}
	\hat{\varPi}_{\bm{n}{'}}^{\dagger\,(-)}.
\end{equation}
Although a direct evaluation is of course possible, it is evident, by construction, that there holds
\begin{equation}
	\langle\underline{\{\varPi\}}\vert\{\varPi{'}\}\rangle=
	\prod_{\bm{n}=-N}^{N}\delta
	\left(\varPi_{0;\,\bm{n}}-\varPi{'}_{0;\,\bm{n}}\right)
	\delta\left(\varPi_{\bm{n}}-\varPi{'}_{\bm{n}}\right)
	\delta\left(\varPi^{\dagger}_{\bm{n}}-\varPi{'}^{\dagger}_{\bm{n}}\right),
\end{equation}
where the last two factors are complex delta functions in general. The completeness of these eigenvectors or the resolution of unity will also be clear because there holds
\begin{equation}
	\int\!\!\prod_{\bm{n}=-N}^{N}
	d\varPi_{0;\,\bm{n}}\,d\varPi_{\bm{n}}\,d\varPi_{\bm{n}}^{\dagger}\,
	\langle\underline{\{\varPhi\}}\vert\{\varPi\}\rangle
	\langle\underline{\{\varPi\}}\vert\{\varPhi{'}\}\rangle=
	\langle\underline{\{\varPhi\}}\vert\{\varPhi{'}\}\rangle.
\end{equation}

We are now ready to formulate path integrals for the model in terms of eigenvectors constructed above. As the frist step, we consider a short time kernel in Euclidean space. To do so, we divide the time interval $-T/2<t<T/2$ into $2N+1$ equal length segments; we also put $T=L$ so that lattice spacing is tuned to be $a$.
By adding source terms, we evaluate
\begin{equation}
\label{eq:kernel01}
	K^{(\mathrm{E})}_{J}[\{\varPhi\},\{\varPhi{'}\};a]=
	\langle\underline{\{\varPhi\}}\vert\left(
	1-a\hat{H}_{0}-
	J_{0;\,\bm{n}}\hat{\phi}_{0;\,\bm{n}}-
	J_{\bm{n}}^{*}\hat{\phi}_{\bm{n}}-
	J_{\bm{n}}\hat{\phi}^{\dagger}_{\bm{n}}\right)
	\vert\{\varPhi{'}\}\rangle,
\end{equation}
by making use of the resolution of unity in terms of the eigenvector $\vert\{\varPi\}\rangle$. Since the Hamiltonian is quadratic, we can perform the Gaussian integrals with respect to $\varPi_{0;\,\bm{n}}$, $\varPi_{\bm{n}}$ and $\varPi^{\dagger}_{\bm{n}}$ immediately to obtain
\begin{equation}
	K^{(\mathrm{E})}_{J}[\{\varPhi\},\{\varPhi{'}\};a]=
	\frac{1}{\sqrt{(2\pi)^{3(2N+1)^{3}}}}
	\exp\left[-\sum_{\bm{n}=-N}^{N}\left\{\frac{1}{2}
	\bm{\varPhi}_{\bm{n}}^{\mathrm{T}}
	\left(-\DAlambert^{(\mathrm{E})}+\bm{M}^{2}\right)\bm{\varPhi}_{\bm{n}}+
	\bm{J}_{\bm{n}}^{\mathrm{T}}\bm{\varPhi}_{\bm{n}}\right\}\right],
\end{equation}
where
\begin{equation}
	\bm{J}_{\bm{n}}^{\mathrm{T}}=\left(J_{0;\,\bm{n}},\,
	J^{*}_{\bm{n}},\,J_{\bm{n}}\right)
\end{equation} 
and we have introduced $\DAlambert^{(\mathrm{E})}$ by
\begin{equation}
	\DAlambert^{(\mathrm{E})}=\tilde{\nabla}_{\mu}\nabla_{\mu}
\end{equation}
in which and hereafter sums with respect to repeated indices are assumed. The rule will also apply even for lattice points as a label of a sum. In the Euclidean formulation, we shall denote $x_{4}=n_{4}a$ and four dimensional lattice points will be labeled simply by $n=(\bm{n},\,n_{4})$. As we will explain soon below, we require the periodic boundary condition also for the time direction. By taking it into account, we have already made use of the integration by parts in
\begin{equation}
	\nabla_{4}\bm{\varPhi}_{\bm{n}}^{\mathrm{T}}
	\nabla_{4}\bm{\varPhi}_{\bm{n}}=-
	\bm{\varPhi}_{\bm{n}}^{\mathrm{T}}\tilde{\nabla}_{4}\nabla_{4}
	\bm{\varPhi}_{\bm{n}}.
\end{equation}

Since our aim is to prepare for the perturbative expansion, we consider here the generating functional $Z^{(\mathrm{E})}[\bm{J}]$ defined by
\begin{equation}
	Z^{(\mathrm{E})}[\bm{J}]=\int\!\!
	\prod_{n=-N}^{N}d\phi_{0;\,n}\,d\phi_{n}\,d\phi^{\dagger}_{n}\,
	K^{(\mathrm{E})}_{J}[\{\varPhi_{N}\},\{\varPhi_{N-1}\};a]
	K^{(\mathrm{E})}_{J}[\{\varPhi_{N-1}\},\{\varPhi_{N-2}\};a]\cdots
	K^{(\mathrm{E})}_{J}[\{\varPhi_{-N}\},\{\varPhi_{N}\};a].
\end{equation}
Here the second argument of the last term above exhibits the periodic boundary condition. If we write the time sliced path integral above as
\begin{equation}
	Z^{(\mathrm{E})}[\bm{J}]=\frac{1}{\sqrt{(2\pi)^{3(2N+1)^{4}}}}
	\int\!\!
	\prod_{n=-N}^{N}d\phi_{0;\,n}\,d\phi_{n}\,d\phi^{\dagger}_{n}\,
	e^{-S^{(\mathrm{E})}_{\bm{J}}[\varPhi]},
\end{equation}
the action of this path integral reads
\begin{equation}
	S^{(\mathrm{E})}_{\bm{J}}[\varPhi]=\frac{1}{2}
	\bm{\varPhi}_{n}^{\mathrm{T}}
	\left(-\DAlambert^{(\mathrm{E})}+\bm{M}^{2}\right)\bm{\varPhi}_{n}+
	\bm{J}_{n}^{\mathrm{T}}\bm{\varPhi}_{n}.
\end{equation}
The action is again, as in the case of quantum mechanical model, separated into three parts corresponding to each degree of freedom; the $\hat{\phi}_{0}$ degree possesses well-defined Euclidean action to yield usual generating functional for a massive scalar field while other degrees have a delicate aspect in the treatment of the zero mode in evaluating functional determinants. Nevertheless, as will be shown in the appendix by the quantum mechanical model, we can obtain well-defined generating functionals for each degree.
We thus obtain $Z^{(\mathrm{E})}[\bm{J}]$ in the factorized form $Z^{(\mathrm{E})}[\bm{J}]=Z^{(\mathrm{E})}_{\phi_{0}}[J_{0}]Z^{(\mathrm{E})}_{\phi}[J^{*}]Z^{(\mathrm{E})}_{\phi^{\dagger}}[J]$;
in this product, each factor being given by
\begin{equation}
	Z^{(\mathrm{E})}_{\phi_{0}}[J_{0}]=
	\frac{1}{\sqrt{\det(-\DAlambert^{(\mathrm{E})}+m^{2})}}
	e^{-W^{(\mathrm{E})}_{\phi_{0}}[J_{0}]},\quad
	W^{(\mathrm{E})}_{\phi_{0}}[J_{0}]=-\frac{1}{2}
	J_{0;\,n}\Delta^{(m)}_{n,\,n{'}}J_{0;\,n{'}},
\end{equation}
\begin{equation}
	Z^{(\mathrm{E})}_{\phi}[J^{*}]=
	\frac{1}{\sqrt{\det(-\DAlambert^{(\mathrm{E})}+iM^{2})}}
	e^{-W^{(\mathrm{E})}_{\phi}[J^{*}]},\quad
	W^{(\mathrm{E})}_{\phi}[J^{*}]=-\frac{1}{2}
	J^{*}_{n}\Delta^{(\omega)}_{n,\,n{'}}J^{*}_{n{'}},
\end{equation}
and
\begin{equation}
	Z^{(\mathrm{E})}_{\phi^{\dagger}}[J]=
	\frac{1}{\sqrt{\det(-\DAlambert^{(\mathrm{E})}-iM^{2})}}
	e^{-W^{(\mathrm{E})}_{\phi^{\dagger}}[J]},\quad
	W^{(\mathrm{E})}_{\phi^{\dagger}}[J]=-\frac{1}{2}
	J_{n}\Delta^{(\omega^{*})}_{n,\,n{'}}J_{n},
\end{equation}
respectively. Here the propagator for each degree is given by the substitution of $m$, $\omega$ and $\omega^{*}$ to $\mu$ in 
\begin{equation}
	\Delta^{(\mu)}_{n,\,n{'}}=\sum_{r_{\nu}=-N}^{N}
	\frac{F_{n_{\nu}}^{r_{\nu}}F_{n{'}_{\nu}}^{r_{\nu}\,*}}
	{4\sum_{\nu=1}^{4}\sin^{2}\left(\dfrac{\pi r_{\nu}}{2N+1}\right)+
	\mu^{2}},\quad
	F^{r_{\nu}}_{n_{\nu}}=
	F^{r_{1}}_{n_{1}}F^{r_{2}}_{n_{2}}F^{r_{3}}_{n_{3}}F^{r_{4}}_{n_{4}}.
\end{equation}
As in the case of quantum mechanical model, we can define the generating functional for the original variable $\hat{\varphi}$ by tuning sources in a harmony to yield
\begin{equation}
	W^{(\mathrm{E})}[J]=-\frac{1}{2}J_{n}\Delta_{n,\,n{'}}J_{n{'}},
\end{equation}
where $\Delta_{n,\,n{'}}$ being given by
\begin{equation}
\begin{aligned}
	\Delta_{n,\,n{'}}=&\frac{1}{\abs{W}^{2}}\Delta^{(m)}_{n,\,n{'}}-
	\frac{1}{2W}\Delta^{(\omega)}_{n,\,n{'}}-
	\frac{1}{2W^{*}}\Delta^{(\omega^{*})}_{n,\,n{'}}\\
	=&
	-M^{4}
	\left\{
	\Delta^{(m)}\Delta^{(\omega)}\Delta^{(\omega^{*})}
	\right\}_{n,\,n{'}}.
\end{aligned}
\end{equation}
It must be stressed here that in the final expression for $\Delta_{n,\,n{'}}$ there appear six poles in total but the way how to avoid these singularities have been already determined in obtaining each propagator from factorized path integral. There is no reason to change the rule just as we like; by changing the rule we will be forced to change boundary conditions for the propagator. Such a change may break the property of the model to fulfill the Euclidicity postulate.

The effect of Lee-Wick ghosts is seen, if we consider a loop integral, which appears in calculation of the self-energy, for example in $V(\varphi)=\lambda\varphi^{4}(x)/4!$ case,
\begin{equation}
	\int_{0}^{\varLambda}\!\!k^{3}dk
	\frac{1}{k^{2}+m^{2}}\frac{M^{4}}{k^{4}+M^{4}}=
	\frac{\pi^{2}M^{4}}{2(m^{4}+M^{4})}\left\{
	2M^{2}\tan^{-1}\frac{\varLambda^{2}}{M^{2}}+m^{2}\log\frac{m^{4}}{M^{4}}+
	\log\frac{\varLambda^{4}+M^{4}}{(\varLambda^{2}+m^{2})^{2}}\right\}.
\end{equation}
Since the right hand side remains finite in the limit $\varLambda\to\infty$, we obtain a finite contribution to the self-energy from this one loop correction.
Thus the Lee-Wick ghosts remove not only the quadratic divergence but also the logarithmic one in this simple loop correction.

Another typical example will be given by
\begin{equation}
	\int\!\!d^{4}k
	\frac{1}{k^{2}+m^{2}}\frac{M^{4}}{k^{4}+M^{4}}
	\frac{1}{(k-p)^{2}+m^{2}}\frac{M^{4}}{\{(k-p)^{2}\}^{2}+M^{4}}
\end{equation}
where $p$ designates the external momentum. By introducing four dimensional polar coordinate, we can carry out integrations with respect to angle variables to obtain a convergent radial integral. In this way, finiteness of the theory can be checked in the Euclidean formulation. Thus our result obtained above can be regarded as the starting point of the perturbative expansion. As we have mentioned above, the contour of the integration with respect to $k_{4}$ is already fixed in the definition of propagators of each component. Our definition of the propagator by the Euclidean path integral gives the reason for the prescription given by Anselmi an Piva in ref.~\cite{Anselmi-Piva}.
In our Euclidean formulation in the present paper, we integrate $k_{4}$ along the real axis in
\begin{equation}
	\Delta^{(\mu)}(x)=
	\int\!\!\frac{d^{4}k}{(2\pi)^{4}}\frac{e^{ik\cdot x}}{k^{2}+\mu^{2}},\quad
	k\cdot x=k_{\mu}x_{\mu}.
\end{equation}
By doing so, we pick up the pole in the lower half-plane for $x_{4}>0$ and the one in the upper half-plane for $x_{4}<0$ to obtain
\begin{equation}
	\Delta^{(\mu)}(x)=
	\int\!\!\frac{d^{3}k}{(2\pi)^{3}2\omega_{\mu}(\bm{k})}
	\left\{\theta(x_{4})e^{-\omega_{\mu}(\bm{k})x_{4}+i\bm{k}\cdot\bm{x}}
	+\theta(-x_{4})e^{\omega_{\mu}(\bm{k})x_{4}-i\bm{k}\cdot\bm{x}}\right\},
\end{equation}
where $\omega_{\mu}(\bm{k})=\sqrt{\bm{k}^{2}+\mu^{2}}$. By this prescription, the propagator is made to remain finite for $\abs{x_{4}}\to\infty$.
To translate it to the propagator in Minkowski space, we just need to put $x_{4}=ix_{0}$ to obtain
\begin{equation}
	\Delta^{(\mu)}_{\mathrm{F}}(x)=
	\int\!\!\frac{d^{3}k}{(2\pi)^{3}2\omega_{\mu}(\bm{k})}
	\left\{\theta(x_{0})e^{-i\omega_{\mu}(\bm{k})x_{0}+i\bm{k}\cdot\bm{x}}
	+\theta(-x_{0})e^{i\omega_{\mu}(\bm{k})x_{0}-i\bm{k}\cdot\bm{x}}\right\}.
\end{equation}
Note that, in this propagator, the contribution from the pole at $k_{0}=\omega_{\mu}(\bm{k})$ is taken into account for $x_{0}>0$ and the one from $k_{0}=-\omega_{\mu}(\bm{k})$ contributes for $x_{0}<0$; that is, the prescription just to put $x_{4}=ix_{0}$ in the Euclidean propagator gives correct method to obtain Lee-Wick prescription. The $S$ matrix for this system is then unitary for any finite time. To define the $S$ matrix, however, for infinite time, we need to introduce the Gaussian adiabatic factor that breaks the Lorentz invariance\cite{Nakanishi:PRD3,Nakanishi:PRD5,Nakanishi:PTPS}. The same holds for the propagator above. If we add $-\epsilon x_{0}^{2}$ to the exponents, we can make the propagator remain finite for $\abs{x_{0}}\to\infty$; but it apparently breaks the Lorentz invariance.

\section{Conclusion}
We have formulated the Euclidean path integral for the scalar LW model. Our method is based on the technique to find eigenvectors of field operators. The technique involves the resolutions of unity, the fundamental ingredient in formulating a time sliced path integral, in itself. Integration contours for ghost degrees are originally defined to be a pair of lines crossing at the origin with each other on the complex plane. It is very interesting to find that, despite the use of indefinite metric representation, we can modify the contours to be along the real axis for ghost degrees in the scalar LW model.
In this regard, the Euclidean path integral of the vector LW model will be the next target for us to construct. This will be reported elsewhere\cite{Sakoda-Suzuki:02}.

Euclidean formulation of path integrals for systems quantized with the indefinite metric representation is the key to define quantum theory for such systems in a proper way. In other words, the Euclidicity postulate\cite{Abers-Lee,Coleman:85} will be the guiding principle to construct path integral representation of such systems. For the scalar LW model, we have successfully formulated the Euclidean path integral and obtained a well-defined propagator to be a basic tool for the perturbative expansion.

\appendix
\section{Treatment of the zero mode in scalar Lee-Wick model}
Let us consider the evaluation of the path integrals in quantum mechanical Lee-Wick model. The degree of $\hat{q}_{0}$ is the well-defined harmonic oscillator and we have nothing to take care about it. Since the other two degrees, on the other hand, possess pure imaginary square frequency, we may encounter a tough trouble in carrying out the "Gaussian" integrals. To make clear this point, we first consider the path integral for $Z^{(\mathrm{E})}_{\phi}[J^{*}]$ here. 

The action of the path integral
\begin{equation}
	Z^{(\mathrm{E})}_{\phi}[J^{*}]=
	\frac{1}{\sqrt{(2\pi\epsilon)^{2N+1}}}\int\!\!\prod_{j=-N}^{N}d\phi_{j}
	e^{-S^{(\mathrm{E})}_{J^{*}}[\phi]}
\end{equation}
reads
\begin{equation}
	S^{(\mathrm{E})}_{J^{*}}[\phi]=\frac{1}{2\epsilon}\phi_{j}
	\left(-\tilde{\nabla}\nabla+iM^{2}\epsilon^{2}\right)\phi_{j}+
	\epsilon J^{*}_{j}\phi_{j}.
\end{equation}
By making a shift $\phi_{j}=\phi^{\mathrm{cl.}}_{j}+\xi_{j}$, we find
\begin{equation}
	S^{(\mathrm{E})}_{J^{*}}[\phi]=
	S^{(\mathrm{E})}_{J^{*}}[\phi^{\mathrm{cl.}}]+
	S^{(\mathrm{E})}_{J^{*}=0}[\xi]+\frac{1}{\epsilon}
	\xi\left\{\left(-\tilde{\nabla}\nabla+iM^{2}\epsilon^{2}\right)
	\phi^{\mathrm{cl.}}_{j}+\epsilon^{2}J^{*}_{j}\right\}.
\end{equation}
We therefore require $\phi^{\mathrm{cl.}}_{j}$ to satisfy
\begin{equation}
	\left(-\tilde{\nabla}\nabla+iM^{2}\epsilon^{2}\right)
	\phi^{\mathrm{cl.}}_{j}+\epsilon^{2}J^{*}_{j}=0.
\end{equation}
Since we can rewrite $S^{(\mathrm{E})}_{J^{*}=0}[\phi^{\mathrm{cl.}}]$ as
\begin{equation}
	S^{(\mathrm{E})}_{J^{*}}[\phi^{\mathrm{cl.}}]=
	\frac{1}{2\epsilon}\phi^{\mathrm{cl.}}_{j}\left\{
	\left(-\tilde{\nabla}\nabla+iM^{2}\epsilon^{2}\right)\phi^{\mathrm{cl.}}_{j}
	+\epsilon^{2}J^{*}_{j}\right\}+
	\frac{\epsilon}{2}J^{*}_{j}\phi^{\mathrm{cl.}}_{j}=
	\frac{\epsilon}{2}J^{*}_{j}\phi^{\mathrm{cl.}}_{j},
\end{equation}
we can evaluate the classical action to be given by
\begin{equation}
	S^{(\mathrm{E})}_{J^{*}}[\phi^{\mathrm{cl.}}]=-\frac{\epsilon^{2}}{2}
	J^{*}_{j}\Delta^{(\omega)}_{j,\,k}J^{*}_{k},
\end{equation}
where we have introduced $\Delta^{(\omega)}_{j,\,k}$ to fulfill
\begin{equation}
	\frac{1}{\epsilon}
	\left(-\tilde{\nabla}\nabla+iM^{2}\epsilon^{2}\right)
	\Delta^{(\omega)}_{j,\,k}=
	\delta_{j,\,k}.
\end{equation}
By assuming the periodic boundary condition, we find that $\Delta^{(\omega)}_{j,\,k}$ is given by
\begin{equation}
	\Delta^{(\omega)}_{j,\,k}=\frac{1}{\epsilon}
	\frac{F_{j}^{n}F_{k}^{n\,*}}
	{\dfrac{4}{\epsilon^{2}}\sin^{2}\left(\dfrac{\pi n}{2N+1}\right)+iM^{2}}.
\end{equation}
(Sum over $n$ is assumed.)
We thus obtain a finite result for $S^{(\mathrm{E})}_{J^{*}}[\phi^{\mathrm{cl.}}]$ without no difficulty though the square frequency is pure imaginary.

Our remaining task is to evaluate the functional determinant arises from integration with respect to $\xi_{j}$s. To achieve this, we introduce the Fourier transform by
\begin{equation}
	\xi_{j}=F_{j}^{n}u_{n},\quad
	u^{*}_{n}=u_{-n},\quad
	u^{*}_{0}=u_{0},
\end{equation}
in particular. Since the Fourier transform is a unitary transformation, the Jacobian in the change of variables from $\xi$s to $u$s must be unity; from the constraint $u^{*}_{n}=u_{-n}$, we can choose $x_{n}$ and $y_{n}$ in $u_{n}=x_{n}+iy_{n}$ for $n=1\,\,2,\,\dots,\,N$ as independent degrees of freedom in addition to the isolated $u_{0}$. The Jacobian of this change of variables yields
\begin{equation}
	\prod_{j=-N}^{N}d\xi_{j}=2^{N}du_{0}\,\prod_{n=1}^{N}dx_{n}\,dy_{n}.
\end{equation}

By rewriting $S^{(\mathrm{E})}_{J^{*}=0}[\xi]$ as
\begin{equation}
	S^{(\mathrm{E})}_{J^{*}=0}[\xi]=\frac{i}{2}M^{2}\epsilon u_{0}^{2}+
	\epsilon\sum_{n=1}^{N}\left\{
	\frac{4}{\epsilon^{2}}\sin^{2}\left(\frac{\pi n}{2N+1}\right)+iM^{2}\right\}
	u_{n}^{*}u_{n},
\end{equation}
we may carry out "Gaussian" integrals with respect to $u_{n}$s. For $n=1,\,2,\,\dots,\,N$, we may rewrite ($\omega^{2}=iM^{2}$)
\begin{equation}
	\epsilon\left\{
	\frac{4}{\epsilon^{2}}\sin^{2}\left(\frac{\pi n}{2N+1}\right)+iM^{2}\right\}
	=\frac{4}{\epsilon}\left\{\sinh^{2}\left(\frac{\omega\epsilon}{2}\right)+
	\sin^{2}\left(\frac{\pi n}{2N+1}\right)\right\}
\end{equation}
to find
\begin{equation}
\begin{aligned}
	&\int\!\!\prod_{n=1}^{N}dx_{n}\,dy_{n}\,
	\exp\left[-\frac{4}{\epsilon}\sum_{n=1}^{N}
	\left\{\sinh^{2}\left(\frac{\omega\epsilon}{2}\right)+
	\sin^{2}\left(\frac{\pi n}{2N+1}\right)\right\}
	\left(x_{n}^{2}+y_{n}^{2}\right)\right]\\
	=&
	\left(\frac{\pi\epsilon}{4}\right)^{N}
	\frac{1}{\displaystyle\prod_{n=1}^{N}
	\left\{\sinh^{2}\left(\dfrac{\omega\epsilon}{2}\right)+
	\sin^{2}\left(\dfrac{\pi n}{2N+1}\right)\right\}}.
\end{aligned}
\end{equation}
Thus, for degrees of $u_{n}$ ($n=1,\,2,\,\dots,\,N$), there appears no problem in carrying out Gaussian integrations.
It is, however, the zero mode ($u_{0}$) that require a careful treatment here. As is evident above, the integrand for this degree is not Gaussian; it is rather a Fresnel integral. This integral can be evaluated by adding a path which goes from real axis down to the line $\Im(u_{0})=-\Re(u_{0})$ then go along this line from lower-right to upper-left and goes down to the negative real axis. We then obtain
\begin{equation}
	\int_{-\infty}^{\infty}\!\!du_{0}e^{-iM^{2}\epsilon u_{0}^{2}/2}=
	\sqrt{\frac{2\pi}{iM^{2}\epsilon}}.
\end{equation}
By making use of a formula
\begin{equation}
	\prod_{n=1}^{N}
	\left\{\sinh^{2}x+\sin^{2}\left(\frac{\pi n}{2N+1}\right)\right\}=
	\frac{\sinh\{(2N+1)x\}}{2^{N}\sinh x},
\end{equation}
we finally obtain
\begin{equation}
	Z^{(\mathrm{E})}_{\phi}[J^{*}]=
	\frac{1}{2\sinh(\omega T/2)}e^{-W^{(\mathrm{E})}_{\phi}[J^{*}]},\quad
	W^{(\mathrm{E})}_{\phi}[J^{*}]=-\frac{\epsilon^{2}}{2}
	J^{*}_{j}\Delta^{(\omega)}_{j,\,k}J^{*}_{k}.
\end{equation}

We can proceed in a quite parallel way and we obtain
\begin{equation}
	Z^{(\mathrm{E})}_{\phi^{\dagger}}[J]=
	\frac{1}{2\sinh(\omega^{*}T/2)}e^{-W^{(\mathrm{E})}_{\phi^{\dagger}}[J]},\quad
	W^{(\mathrm{E})}_{\phi^{\dagger}}[J]=-\frac{\epsilon^{2}}{2}
	J_{j}\Delta^{(\omega^{*})}_{j,\,k}J_{k}.
\end{equation}
Note that the Fresnel integration for the zero mode for this case requires integration along a line which is orthogonal to the previous case; although the difference is harmless in the Euclidean formulation, it seems to break the convergence of the path integral in the former case if we consider the corresponding path integrals in the Minkowski space (real time). This observation is, however, wrong and we can define the Minkowski generating functionals for both systems if we make use of the coherent states.
We will show this shortly in the following.

For the quantum mechanical Lee-Wick model, we may express the free Hamiltonian in terms of the creation and annihilation operators to find
\begin{equation}
	\hat{H}_{0}=m\left(\hat{a}_{0}^{\dagger}\hat{a}_{0}+1/2\right)+
	\omega\left(\hat{\beta}^{\dagger}\hat{\alpha}+1/2\right)+
	\omega^{*}\left(\hat{\alpha}^{\dagger}\hat{\beta}+1/2\right).
\end{equation}
We take the last two terms as the Hamiltonian $\hat{H}{'}$ in the following. By introducing source terms as $f^{*}\hat{\alpha}+f\hat{\beta}^{\dagger}+g^{*}\hat{\beta}+g\hat{\alpha}^{\dagger}$ at each lattice point in the time sliced path integral, we consider the trace formula:
\begin{multline}
	Z^{(\mathrm{E})}[f,g]=\int\!\!\prod_{j=-N}^{N}
	\frac{d\alpha_{j}^{*}\,d\alpha_{j}\,d\beta_{j}^{*}\,d\beta_{j}}{\pi^{2}}
	\exp\left[-\sum_{j=-N}^{N}\left\{
	\alpha^{*}_{j}\alpha_{j}+\beta^{*}\beta_{j}-
	e^{-\omega\epsilon}\alpha^{*}_{j}\alpha_{j-1}-
	e^{-\omega^{*}\epsilon}\beta^{*}_{j}\beta_{j-1}\right.\right.\\
	+\left.\left.\vphantom{\sum_{j=-N}^{N}}\left.
	\epsilon\left(f^{*}_{j}\alpha_{j-1}+f_{j}\alpha^{*}_{j}+
	g^{*}_{j}\beta_{j-1}+g_{j}\beta^{*}_{j}\right)\right\}
	-\frac{1}{2}(\omega+\omega^{*})T\right]\right\vert_{\mathrm{P.B.C}},
\end{multline}
where P.B.C denotes the periodic boundary condition. Clearly the path integral is factorized into two factors. We therefore consider them separately.

First, we consider
\begin{equation}
	Z^{(\mathrm{E})}_{(\omega)}[f]=
	\int\!\!\prod_{j=-N}^{N}\frac{d\alpha_{j}^{*}\,d\alpha_{j}}{\pi}\,
	\exp\left[-\sum_{j=-N}^{N}\left\{\alpha^{*}_{j}\left(
	\alpha_{j}-e^{-\omega\epsilon}\alpha_{j-1}\right)+
	\epsilon\left(f^{*}_{j}\alpha_{j-1}+f_{j}\alpha^{*}_{j}\right)\right\}-
	\frac{1}{2}\omega T\right].
\end{equation}
We write the Euclidean action in the exponent above as $S^{(\mathrm{E})}_{(\omega);\,f}[\alpha,\,\alpha^{*}]$ and rewrite it as
\begin{equation}
	S^{(\mathrm{E})}_{(\omega);\,f}[\alpha,\,\alpha^{*}]=\bm{\alpha}^{\dagger}
	\left(\bm{1}-e^{-\omega\epsilon}\varOmega\right)\bm{\alpha}+
	\epsilon\bm{f}^{\dagger}\varOmega\bm{\alpha}+
	\epsilon\bm{\alpha}^{\dagger}\bm{f},
\end{equation}
where we have put
\begin{equation}
	\bm{\alpha}^{\dagger}=\left(\alpha^{*}_{-N},\,\alpha^{*}_{-N+1},\,\dots,\,
	\alpha^{*}_{N-1},\,\alpha^{*}_{N}\right),\quad
	\bm{f}^{\dagger}=\left(f^{*}_{-N},\,f^{*}_{-N+1},\,\dots,\,
	f^{*}_{N-1},\,f^{*}_{N}\right),
\end{equation}
etc., and defined a $(2N+1)\times(2N+1)$ matrix $\varOmega$ whose components being given by $\varOmega_{j,\,k}=\delta_{j,\,k+1}+\delta_{N-k+j,\,-N}$ where $j$ and $k$ runs from $-N$ through $N$ and there holds $\varOmega^{2N+1}=\bm{1}$, where and in the above $\bm{1}$ is $(2N+1)\times(2N+1)$ unit matrix.
The determinant and the inverse of the matrix $\bm{1}-e^{-\omega\epsilon}\varOmega$ above is found in a straightforward manner; by the use of cofactor expansion
\begin{equation}
	\det(\bm{1}-e^{-\omega\epsilon}\varOmega)=
	\begin{vmatrix}
	1&0&0&\cdots&\cdots&0&-\gamma\\
	-\gamma&1&0&0&\cdots&\cdots&0\\
	0&-\gamma&1&\ddots&\ddots&&\vdots\\
	0&0&-\gamma&1&\ddots&\ddots&\vdots\\
	\vdots&&\ddots&\ddots&\ddots&\ddots&0\\
	\vdots&&&&\ddots&\ddots&0\\
	0&\cdots&\cdots&\cdots&0&-\gamma&1
	\end{vmatrix}=1-\gamma^{2N+1},\quad
	\gamma=e^{-\omega\epsilon},
\end{equation}
and from
\begin{equation}
	\left(\bm{1}-\gamma\varOmega\right)
	\left(\bm{1}+\gamma\varOmega+\gamma^{2}\varOmega^{2}+\cdots+
	\gamma^{2N-1}\varOmega^{2N-1}+\gamma^{2N}\varOmega^{2N}\right)
	=1-\gamma^{2N+1}
\end{equation}
we obtain, for $\gamma^{2N+1}\ne1$,
\begin{equation}
	\left(\bm{1}-\gamma\varOmega\right)^{-1}=
	\frac{1}{1-\gamma^{2N+1}}
	\left(\bm{1}+\gamma\varOmega+\gamma^{2}\varOmega^{2}+\cdots+
	\gamma^{2N-1}\varOmega^{2N-1}+\gamma^{2N}\varOmega^{2N}\right).
\end{equation}
For the present case $\gamma=e^{-\omega\epsilon}$ and thus $\gamma^{2N+1}=e^{-\omega T}$. As commented in the text, $\omega$ possesses positive real part so that we can claim $\gamma^{2N+1}\ne1$.
It is important here to note that there is no difficulty in the process of finding $\det(\bm{1}-e^{-\omega\epsilon}\varOmega)$ and $(\bm{1}-e^{-\omega\epsilon}\varOmega)^{-1}$ above. It remains true even if we replace $T$ by $iT$. (For that case, due to the positive imaginary part of $\omega$, we can claim again $\gamma^{2N+1}\ne1$ as far as $T>0$.)

By carrying out Gaussian integrals after completing the square, we obtain
\begin{equation}
	Z^{(\mathrm{E})}_{(\omega)}[f]=
	\frac{1}{2\sinh(\omega T/2)}
	e^{-W^{(\mathrm{E})}_{(\omega)}[f]},\quad
	W^{(\mathrm{E})}_{(\omega)}[f]=-
	\epsilon^{2}\bm{f}^{\dagger}\varOmega
	(\bm{1}-e^{-\omega\epsilon}\varOmega)^{-1}
	\bm{f}.
\end{equation}
Since we know the components of the matrix $(\bm{1}-e^{-\omega\epsilon}\varOmega)^{-1}$ are given by
\begin{equation}
	(\bm{1}-e^{-\omega\epsilon}\varOmega)^{-1}_{j,\,k}=
	\frac{1}{2\sinh(\omega T/2)}\left\{
	\theta(j-k)e^{\omega T/2-\omega(j-k)\epsilon}+
	\theta(k-j)e^{-\omega T/2+(k-j)\epsilon}\right\},
\end{equation}
where $\theta(j-k)$ is the discrete step function defined by
\begin{equation}
	\theta(j-k)=\begin{cases}
	1&(j\ge k)\\
	&\\
	0&(j<k)
	\end{cases},
\end{equation}
we can express the generating functional $W^{(\mathrm{E})}_{(\omega)}[f]$ as
\begin{equation}
	W^{(\mathrm{E})}_{(\omega)}[f]=-\epsilon^{2}\sum_{j,\,k=-N}^{N}
	f^{*}_{j+1}\frac{1}{2\sinh(\omega T/2)}\left\{
	\theta(j-k)e^{\omega T/2-\omega(j-k)\epsilon}+
	\theta(k-j)e^{-\omega T/2+\omega(k-j)\epsilon}\right\}f_{k},
\end{equation}
in which, due to the periodic boundary condition, $f^{*}_{N+1}=f^{*}_{-N}$.

Next, we should consider
\begin{equation}
	Z^{(\mathrm{E})}_{(\omega^{*})}[g]=
	\int\!\!\prod_{j=-N}^{N}\frac{d\beta_{j}^{*}\,d\beta_{j}}{\pi}\,
	\exp\left[-\sum_{j=-N}^{N}\left\{\beta^{*}_{j}\left(
	\beta{j}-e^{-\omega^{*}\epsilon}\beta{j-1}\right)+
	\epsilon\left(g^{*}_{j}\beta{j-1}+g_{j}\beta^{*}_{j}\right)\right\}-
	\frac{1}{2}\omega^{*}T\right]
\end{equation}
but the evaluation of this generating functional is completely parallel to the one above. We thus omit the derivation and show here the result:
\begin{equation}
\begin{aligned}
	Z^{(\mathrm{E})}_{(\omega^{*})}[g]=&
	\frac{1}{2\sinh(\omega^{*}T/2)}
	e^{-W^{(\mathrm{E})}_{(\omega^{*})}[g]},\\
	W^{(\mathrm{E})}_{(\omega^{*})}[g]=&
	-\epsilon^{2}\sum_{j,\,k=-N}^{N}
	g^{*}_{j+1}\frac{1}{2\sinh(\omega^{*}T/2)}\left\{
	\theta(j-k)e^{\omega^{*}T/2-\omega^{*}(j-k)\epsilon}+
	\theta(k-j)e^{-\omega^{*}T/2+\omega^{*}(k-j)\epsilon}\right\}g_{k}.
\end{aligned}
\end{equation}

We now put $f^{*}_{j}=f_{j}=J^{*}_{j}/\sqrt{2\omega}$ and $g^{*}_{j}=g_{j}=J_{j}/\sqrt{2\omega^{*}}$ to fit the sources in the text and discard higher order powers of $\omega\epsilon$ and $\omega^{*}\epsilon$ to obtain
\begin{equation}
	W^{(\mathrm{E})}_{(\omega)}[J^{*}]=-\frac{\epsilon^{2}}{2}
	\sum_{j,\,k=-N}^{N}
	J^{*}_{j}\Delta^{(\omega){'}}_{j,\,k}J^{*}_{k},\quad
	\Delta^{(\omega){'}}_{j,\,k}=
	\frac{1}{2\omega\sinh(\omega T/2)}
	\cosh\left\{\omega\left({\frac{T}{2}-\abs{j-k}\epsilon}\right)\right\},
\end{equation}
and
\begin{equation}
	W^{(\mathrm{E})}_{(\omega^{*})}[J]=
	-\frac{\epsilon^{2}}{2}
	\sum_{j,\,k=-N}^{N}
	J_{j}\Delta^{(\omega^{*}){'}}_{j,\,k}J_{k},\quad
	\Delta^{(\omega^{*}){'}}_{j,\,k}=
	\frac{1}{2\omega^{*}\sinh(\omega^{*}T/2)}
	\cosh\left\{\omega^{*}\left({\frac{T}{2}-\abs{j-k}\epsilon}\right)\right\},
\end{equation}
where propagators $\Delta^{(\omega){'}}$ and $\Delta^{(\omega^{*}){'}}$ above are, in the continuum limit, equivalent to the ones $\Delta^{(\omega)}$ and $\Delta^{(\omega^{*})}$ in the text and their behavior in the limit $T\to\infty$ read
\begin{equation}
	\Delta^{(\omega){'}}\sim
	\frac{1}{2\omega}e^{-\abs{j-k}\omega\epsilon},\quad
	\Delta^{(\omega^{*}){'}}\sim
	\frac{1}{2\omega^{*}}e^{-\abs{j-k}\omega^{*}\epsilon}.
\end{equation}
We thus observe that in the Euclidean formulation the behavior of both propagators of $\hat{\phi}$ and $\hat{\phi}^{\dagger}$ are finite though it changes in the Minkowski (real) time.

\end{document}